\documentstyle[preprint,aps,eqsecnum]{revtex}
\tighten
\begin{document}
\def\eps{\varepsilon}
\def\klein#1{{\mbox{\scriptsize #1}}}
\def\pzero{{\phantom{0}}}

\draft

\title{Magnetoconductance of ballistic chaotic quantum dots: \\
A Brownian motion approach for the $S$-matrix.}

\author{Klaus Frahm and Jean-Louis Pichard}

\address{Service de Physique de l'\'Etat condens\'e.
	CEA Saclay, 91191 Gif-sur-Yvette, France.}

\date{\today}

\maketitle

\begin{abstract}
  Using the Fokker-Planck equation describing the evolution of the
transmission eigenvalues for Dyson's Brownian motion ensemble,
we calculate the magnetoconductance of a ballistic chaotic dot in
in the crossover regime from the orthogonal to the unitary symmetry.
The correlation functions of the transmission eigenvalues are
expressed in terms of quaternion determinants for arbitrary number
$N$ of scattering channels. The corresponding average, variance
and autocorrelation function of the magnetoconductance are
given as a function of the Brownian motion time $t$. A microscopic
derivation of this $S$-Brownian motion approach is discussed and
$t$ is related to the applied flux. This exactly solvable random
matrix model yields the right expression for the suppression of the
weak localization corrections in the large $N$-limit and for small applied
fluxes. An appropriate rescaling of $t$ could extend its validity
to larger magnetic fluxes for the averages, but not for the correlation
functions.
\end{abstract}

\tableofcontents

\section{Introduction}

\label{section:1}

 Ballistic transport through quantum dots of various shapes are
the subject of many theoretical and experimental investigations.
One of the fundamental issues is to determine in a conductance measurement
the signature of ``quantum chaos'', i. e. of the quantum analog
of a classically chaotic dynamics. Usually, such dots are made
with two dimensional semiconductor nanostructures (quantum billiards)
connected to two electron reservoirs,
and one measures the magnetic field dependence of the conductance
at very low temperature\cite{marcus,keller,chang}
(quantum coherent transport).
The question is to describe the statistical properties of the
recorded magneto-fingerprints for different shapes of the dot.
We restrict ourselves to fully chaotic systems where we hope to
state universal results, leaving aside the integrable systems.
The usual theoretical approaches\cite{berry1,jalabert3}
concentrate on microscopic models, using numerical calculations
and semiclassical approximations, or more or less straightforward
random matrix models, possibly combined with supersymmetric methods.
The semiclassical approach requires in the final evaluation the
so-called diagonal approximation,
which is problematic \cite{jalabert3} for the total weak localization
correction. Supersymmetry\cite{efetov,verbaarschot} and random matrix
model have also been recently used \cite{pluhar} to describe the
ensemble averaged behavior of the magneto conductance. In Ref. \cite{pluhar}
the Hamiltonian
of the dot was modeled by a Pandey-Mehta random matrix Hamiltonian
\cite{pandey1}, but the used method
 leads to substantial technical problems and the averaged magneto conductance
is expressed in terms of a rather complicated integral expression which
can be evaluated by numerical means.

Another random matrix approach for the quantum billiards
was used in Refs. \cite{jalabert1,baranger1} and models directly
the scattering matrix $S$ of the dot instead of its Hamiltonian.
There are reasons \cite{bohigas2,bluemel} to assume that $S$
is suitably described by one of Dyson's classical circular ensembles
\cite{dyson1} (COE-CUE-CSE). In
Refs. \cite{jalabert1,baranger1}, the associated transport properties
have been calculated: e. g. for the circular orthogonal ensemble (COE)
and for circular unitary ensemble (CUE) corresponding to chaotic dots
without or with a large enough applied magnetic field respectively.
We propose a
generalization of this approach in terms of a continuous crossover
from COE to CUE, that is described by one of Dyson's Brownian motion
ensembles \cite{dyson2}. The idea is to consider a Brownian motion with
a fictitious time $t$ that leads to a {\em diffusion\/} in $S$-matrix space.
In the limit $t\to\infty$ the probability distribution diffuses to a
stationary solution that corresponds in our application to the circular
unitary ensemble. Choosing as initial condition for $t=0$ the circular
orthogonal ensemble, we have a model for the magnetoconductance of a
chaotic dot, at least as a function of the Brownian motion time $t$.
Before completion of this manuscript, we received a preprint by
Jochen Rau \cite{rau} where a very qualitative analysis of this model
is presented with essentially similar conclusions than our detailed
study. One can also mention that this crossover ensemble was already
considered in Refs. \cite{shukla1,shukla2} where the correlation
functions for the scattering phase shifts $\theta_i$ were calculated.
The decisive difference here is that we are interested in the transport
properties, i. e. in the statistics of the transmission eigenvalues
$T_i$ of $tt^\dagger$, when $S$ is parametrized by
\begin{equation}
\label{smatrix}
S=\left(\begin{array}{cc}
r & t' \\
t & r' \\
\end{array}
\right)\quad.
\end{equation}
The Brownian motion leads to a Fokker-Planck equation for the probability
distribution of the variables $T_i$ which have been first derived in
Ref. \cite{frahm1}. This Fokker-Planck equation can be mapped onto a
Schr\"odinger equation of a one dimensional system of free Fermions
and solved for an arbitrary initial condition. The
corresponding propagator was already calculated in Ref. \cite{frahm1}
where the crossover of two decoupled CUEs of dimension $N$ to one
CUE of dimension $2N$ was considered. Here we use another initial
condition that is just the joint probability distribution for
the variables $T_i$ of the COE-case. The time dependent solution of the
Fokker-Planck equation is then the joint probability distribution for the
crossover ensemble COE $\to$ CUE.

 This paper is subdivided in two main parts (Sections
\ref{section:2} and \ref{section:3}). Section \ref{section:2} is
completely devoted to an exact solution of the Brownian motion
model as it stands. We give any $k$-point correlation
function for the $T_i$ at one particular time $t$. In addition,
we give also the most simple correlation functions
between two different time values. The technical method to obtain
these results is based on a formulation in terms of quaternion
determinants \cite{dyson4} and skew orthogonal polynomials.
The final results for the correlation functions are given
in terms of Legendre Polynomials and can be easily evaluated by numerical
means. The averages for the conductance and the autocorrelation of
its variance (between different Brownian motion times) are exactly
calculated for an arbitrary value of the channel number.
All results of Section \ref{section:2} are exact consequences of
the $S$-Brownian motion ensemble applied to the COE $\to$ CUE crossover.

 It remains to express the Brownian motion time $t$ in terms of a real
physical parameter such as the applied magnetic flux $\Phi$. This issue
is considered in detail in Section \ref{section:3}.
We first discuss the relation of the Pandey-Mehta
Hamiltonian \cite{pandey1} with finite disordered systems
(in the diffusive regime) or with quantum billiards (in the
chaotic ballistic regime). Both cases have already been
considered in Refs. \cite{dupuis1,bohigas4,bohigas5}
and we give a simple treatment that accounts for both cases and
generalizes the known results to the quantum billiard case. In
subsection \ref{section:2b} we briefly recall the different
magnetic field scales in a quantum billiard and the effects of the
Landau quantization for very strong fields.  In subsection
\ref{section:3c}, a microscopic justification\footnote{
We mention that in Ref. \cite{shukla2} a semiclassical and numerical
justification for the validity of the Brownian motion ensemble
has been given as far as the scattering phase shifts correlations
are concerned.} of the Brownian motion
ensemble is sketched, assuming a crossover Pandey-Mehta Hamiltonian
for $H$ and using a well known expression of $S$ in terms
of $H$. We obtain
an explicit expression that relates the Brownian motion time
with the Pandey-Mehta parameter $\alpha$ and therefore with the
magnetic flux. Using this translation we find a Gaussian
instead of a Lorentzian behavior (as given in Ref. \cite{pluhar})
for the suppression of the weak localization correction to the
averaged conductance, when a flux $\Phi$ is applied. One finds
a certain characteristic flux $\Phi_C$ separating two regimes.
For a small flux $\Phi \le \Phi_C$ our result matches precisely
that of Ref. \cite{pluhar}. For a large flux $\Phi\gg \Phi_C$ the
discrepancy between the Gaussian and Lorentzian becomes more
important. This indicates that the Brownian motion
model is not precisely equivalent to the model used in \cite{pluhar}
for large times of the Brownian motion. The difficulty to obtain
directly from our model the lorentzian shape for the weak localization
suppression and the lorentzian {\it square} behavior for the correlation
functions is discussed in conclusion.

\section{$S$-Brownian motion ensemble and COE to CUE crossover}

\label{section:2}

In Ref. \cite{frahm1} we used the original Dyson Brownian motion
ensemble for the scattering matrix to obtain a Fokker-Planck
equation for the {\it transmission eigenvalues} $T_i$.
This is suitable for having the conductance $g$ of a scatterer
connected by $N$-channel contacts to the electron reservoirs,
since $g =\sum_{i=1}^N T_i$.
The stationary solution of the Fokker-Planck equation is given
by the transmission eigenvalue distribution of the circular unitary
ensemble describing a chaotic dot without time reversal
invariance. The chosen initial condition for the Brownian motion is
the COE-distribution for the $T_i$. Concerning the definition
of the Brownian motion,
the derivation of the Fokker-Planck equation for the transmission eigenvalues
and its solution for an arbitrary initial condition,  we refer the reader
to Ref. \cite{frahm1}. It is convenient to use the
same coordinates $x_i=1-2T_i$ of \cite{frahm1}, where the $T_i$ are the
eigenvalues of $tt^\dagger$.

\subsection{Joint probability distribution of the transmission eigenvalues}

\label{section:2a}

The joint probability distribution $p(x,t)$ for the Brownian
motion ensemble can be expressed \cite{frahm1} as follows
\begin{equation}
\label{xsol1}
p(x,t)=\int d^N\tilde x\ \hat p(\tilde x)
\,G(\tilde x,x;t)\quad,
\end{equation}
where the many particle Green's function is given by
\begin{eqnarray}
\label{xgreen0}
G(\tilde x,x;t) & = & \frac{1}{N!} \rho(\tilde x)^{-1}
\ \rho(x)\ \det(g(x_i,\tilde x_j;t))\ e^{-C_N t}\quad,\\
\label{xrhodef}
\rho(x) & = & \prod_{i>j}^N \left(x_i-x_j\right)\quad,\\
\label{xgreen1}
g(x_i,\tilde x_j;t) & = & \sum_{n=0}^\infty
\frac{1}{2}(1+2n)\,P_n(x_i)\,P_n(\tilde x_j)\ e^{-\eps_n t}\quad.
\end{eqnarray}
$P_n(x_i)$ is the Legendre polynomial of degree $n$, $\eps_n=(1+2n)^2$ is the
eigenvalue of the ``one particle Hamiltonian'' \cite{frahm1} and the
constant $C_N$ is given by the sum $C_N=-\sum_{n=0}^{N-1} \eps_n$.
The function $\hat p(x)=p(x,0)$ is the initial condition. In our case
it is given by the joint probability distribution for the orthogonal
circular ensemble which in terms of the variables $x_j$ reads as
\cite{jalabert1,baranger1}
\begin{equation}
\label{prob_orth}
\hat p(x) \propto \prod_{i=1}^N \frac{1}{\sqrt{2(1-x_i)}}\,
\prod_{i>j}^N \left|x_i-x_j\right|\quad.
\end{equation}
The integral in Eq. (\ref{xsol1}) can be evaluated by the method of
integration over alternate variables \cite{mehta}. As usually in such
situations, we consider now the case of an even number $N$ of channels. The
complications that arise from the case of odd $N$ will be discussed
in Appendix \ref{append:b}. Our final results of this Section
are valid for all values of $N$. The result for the joint probability
distribution can be expressed in terms of the so-called Pfaffian
\begin{equation}
\label{prob_pfaff}
p(x,t) \propto \prod_{i>j}^N \left|x_i-x_j\right|
\sqrt{\det(H(x_i,x_j;t)}\ e^{-C_N t}
\end{equation}
where $H(x_i,x_j;t)$ is an antisymmetric function given by
the double integral
\begin{equation}
\label{h_def}
H(x_1,x_2;t)=\frac{1}{2}\int_{-1}^1 dy_1\int_{-1}^1 dy_2
\ \frac{\eps(y_2-y_1)}{\sqrt{(1-y_2)(1-y_1)}}
\ g(x_1,y_1;t)\,g(x_2,y_2;t)
\end{equation}
where $\eps(u)=+1,\ 0,\ -1$ according to the cases $u>0,\ =0,\ <0$.
In the next subsection we will present a formulation in terms
of quaternion determinants \cite{dyson3} that enables us to evaluate
all types of correlation functions and to get exact expressions for
the average conductance and the conductance fluctuations.

 We first briefly discuss the implications of Eq. (\ref{prob_pfaff}) for
the ``level'' repulsion in a more qualitative way. For this
discussion the angles $\varphi_j$ with $x_j=\sin^2\varphi_j$ are
more appropriate. The one particle Green's function $g(x_i,x_j;t)$
fulfills in terms of the $\varphi_j$ a time dependent
Schr\"odinger equation \cite{frahm1} with a particular potential.
In the (extreme) short time limit one can neglect the effect of this
potential and the Green's function can be approximated by the
free ``diffusion propagator'', i.e.
\begin{equation}
\label{free_prop}
g(\sin^2(\varphi_i),\sin^2(\varphi_j);t)\simeq
\frac{1}{\sqrt{\sin(2\varphi_i)\sin(2\varphi_j) 16 \pi t}}
\ \exp\left(-\frac{1}{4t}(\varphi_i-\varphi_j)^2\right)\quad.
\end{equation}
Using this expression we can evaluate the
integral in Eq. (\ref{h_def}) and we obtain for the joint probability
distribution for the variables $\varphi_i$ in the limit $t\to 0$
the expression
\begin{equation}
\label{prob_phi_approx}
\tilde p(\varphi,t)\simeq F_1(\varphi) \sqrt{\det\left(\mbox{erf}\left(
(\varphi_i-\varphi_j)/\sqrt{4t}\,
\right)\right)}
\end{equation}
where $F_1(\varphi)$ is the joint probability distribution for the
orthogonal case, i.e. for exactly $t=0$. Pandey and Mehta \cite{pandey1}
obtained for the eigenvalue distribution of the random matrix
Hamiltonian described by Eq. (\ref{pm_hamiltonian}) a very similar
expression if the $\varphi_j$ are identified with the energy levels
of $E_\alpha$ and if one relates the symmetry breaking
parameters $\alpha$ and $t$ in a suitable way. The expression
(\ref{prob_phi_approx}) shows that the crossover from COE to CUE
has an effect for arbitrarily small values of $t>0$ if the difference
$|\varphi_i-\varphi_j|$ is sufficiently small. Let us first
consider the situation where all $\varphi_i$ are well separated, i.e.
$|\varphi_i-\varphi_j|\gg \sqrt{4t}$. In this case the error function
in Eq. (\ref{prob_phi_approx}) can be replaced by
$\mbox{sign}(\varphi_i-\varphi_j)$ and the determinant is just one.
The probability distribution then equals the distribution $F_1(\varphi)$
for the COE. On the other hand the determinant that appears in
(\ref{prob_phi_approx}) is an even function of
$\varphi_i-\varphi_j$ and vanishes if one of these differences is zero.
If for example two values of the $\varphi_i$ are very close, e.g.
$|\varphi_1-\varphi_2|\ll \sqrt{4t}$, the square root of the determinant
becomes proportional to the difference $|\varphi_1-\varphi_2|/\sqrt{4t}$
which leads together with the factor $F_1(\varphi)$ to the {\it quadratic}
level repulsion of the unitary case. This behavior means that the COE
to CUE crossover for the transmission eigenvalue correlations appears
at arbitrarily
small values for the time parameter $t$ on scales smaller than $\sqrt{4t}$.
The same effect is also observed for the energy levels $E_\alpha$
\cite{pandey1} and for the scattering phase shifts $\theta_i$ \cite{shukla1}.

\subsection{Quaternion formulation}

\label{section:2b}

In this subsection, we use the technic of quaternions, that
was introduced in the theory of random matrices in Ref. \cite{dyson3},
to calculate the $m$-point correlation functions
given by the usual definition
\begin{equation}
\label{rm_def}
R_m(x_1,\ldots,x_m;t)=\frac{N!}{(N-m)!}\int_{-1}^1 dx_{m+1}
\cdots\int_{-1}^1 dx_N\ p(x,t)\quad.
\end{equation}

We define for two
arbitrary functions $f$, $g$ the antisymmetric scalar product
\begin{equation}
\label{skal_def}
<f,g>^{(t)}=\int_{-1}^1 dy_1\int_{-1}^1 dy_2
\ H(y_1,y_2;t)\,f(y_1)\,g(y_2)
\end{equation}
where $H(y_1,y_2;t)$ is just given by Eq. (\ref{h_def}). Suppose
that the number $N$ of channels is even and that we have a set
of skew orthogonal polynominals $q_n^{(t)}(x)$
of degree $n=0,\ldots,N-1$ with
respect to the scalar product (\ref{skal_def}), i.e.
\begin{equation}
\label{skew_orth}
<q_n^{(t)},q_m^{(t)}>^{(t)}=Z_{nm}
\end{equation}
where $Z_{nm}$ is an antisymmetric matrix with $Z_{2n,2n+1}=-Z_{2n+1,2n}=1$
and all other $Z_{nm}=0$. The superscript means that both the
polynomials and the scalar product depend explicitly on the time
parameter $t$. At the moment, we do not consider explicit expressions
for the $q_n^{(t)}(x)$ and the following method remains
general.
In addition, we introduce the functions $Q_n^{(t)}(x)$ via
\begin{equation}
\label{dual_def}
Q_n^{(t)}(y_2)=\int_{-1}^1 dy_1\,H(y_1,y_2;t)\ q_n^{(t)}(y_1)
\end{equation}
which are dual to the $q_n^{(t)}(x)$ :
\begin{equation}
\label{dual_orth}
\int_{-1}^1 dy\ Q_n^{(t)}(y)\,q_m^{(t)}(y)=Z_{nm}\quad.
\end{equation}
We define furthermore functions of two variables $x$, $y$ by
\begin{equation}
\label{kk_def}
K_{F\tilde F}(x,y;t)=\sum_{n,m=0}^{N-1} F_n^{(t)}(x)\,Z_{nm}\,
\tilde F_m^{(t)}(y)
\end{equation}
where $F$ and $\tilde F$ stand for one of the symbols $q$ or $Q$.
Hence, Eq. (\ref{kk_def}) defines four different functions $K_{qq}$,
$K_{qQ}$, $K_{Qq}$ and $K_{QQ}$. From the definition of the matrix $Z_{nm}$
and Eqs. (\ref{dual_def}-\ref{kk_def})
we can immediately state the following properties
\begin{eqnarray}
\label{kk_prop1}
K_{F\tilde F}(x,y;t) & = & -K_{\tilde F F}(y,x;t)\quad,\\
\label{kk_prop2}
-\int_{-1}^1 dx\ K_{qQ}(x,x;t) & = & \int_{-1}^1 dx\ K_{Qq}(x,x;t)=N\quad,\\
\label{kk_prop3}
\int_{-1}^1 dy\ K_{Fq}(x,y;t)\,K_{Q\tilde F}(y,z) & = &
-\int_{-1}^1 dy\ K_{FQ}(x,y;t)\,K_{q\tilde F}(y,z)=
K_{F\tilde F}(x,z;t)\quad,
\end{eqnarray}
and
\begin{equation}
\label{kk_prop4}
\int_{-1}^1 dy\ H(y,x;t)\,K_{qF}(y,z;t)=
K_{QF}(x,z;t).
\end{equation}
 From now on, we strongly refer to the notations and results of Ref.
\cite{dyson3} concerning quaternion properties.
We introduce the $2\times 2$-matrix
(or quaternion) function $\sigma(x,y;t)$ by
\begin{equation}
\label{sig_def}
\sigma(x,y;t)=\left(
\begin{array}{cc}
-K_{qQ}(x,y;t)  &  K_{qq}(x,y;t) \\
H(x,y;t)-K_{QQ}(x,y;t) &  K_{Qq}(x,y;t) \\
\end{array}
\right)
\end{equation}
that fulfills the three properties
\begin{eqnarray}
\label{sig_prop1}
\int_{-1}^1 dx\ \sigma(x,x;t)&=&N
\left(\begin{array}{cc}
1 & 0 \\
0 & 1 \\
\end{array}\right)\quad\\
\label{sig_prop2}
\int_{-1}^1 dy\ \sigma(x,y;t)\,\sigma(y,z;t)&=&
\sigma(x,z;t)+\tau\,\sigma(x,z;t)-\sigma(x,z;t)\,\tau\quad,\quad\\
p(x,t)&=&const.\cdot\mbox{QDet}\left((\sigma(x_i,x_j)_{1\le i,j\le N}\right)
\label{sig_prop3}
\end{eqnarray}
where $\tau=\frac{1}{2}{1\ \phantom{-}0
\choose 0\ -1}$ and QDet$(\cdots)$ is the quaternion-determinant defined
in Ref. \cite{dyson3}. The verification of Eqs. (\ref{sig_prop1}) and
(\ref{sig_prop2}) is a straightforward calculation applying the
properties (\ref{kk_prop1}-\ref{kk_prop4}). The proof of
Eq. (\ref{sig_prop3}) is not so obvious and can be obtained as follows.
First, we note that (\ref{kk_prop1}) and the antisymmetry
of $H(x,y;t)$ imply that the quaternion-matrix $\sigma(x_i,x_j;t)$
is self-dual, i.e.
\begin{equation}
\label{self_dual}
\sigma(x_i,x_j;t)=\overline{\sigma}(x_j,x_i;t)
\end{equation}
where
\begin{equation}
\label{q_adjoint}
\overline{
\left(
\begin{array}{cc}
a & b \\
c & d \\
\end{array}
\right)}
=\left(
\begin{array}{cc}
d & -b \\
-c & a \\
\end{array}
\right)
\end{equation}
is the quaternion adjoint in the notation of Ref. \cite{dyson3}. As in
Ref. \cite{dyson3} we denote with $A(\sigma(x_i,x_j;t))$ the conventional
$2N\times 2N$-matrix that contains the quaternions as $2\times 2$ blocks.
{}From theorem 2 of \cite{dyson3} we get the identity (QDet$(\sigma(x_i,x_j;t
)))^2=\det(A(\sigma(x_i,x_j;t)))$ that relates the conventional with the
quaternion-determinant for self-dual matrices. We consider therefore
the matrix
\begin{eqnarray}
\nonumber
A(\sigma(x_i,x_j;t))&=&\left(
\begin{array}{cc}
-K_{qQ}(x_i,x_j;t)  &  K_{qq}(x_i,x_j;t) \\
H(x_i,x_j;t)-K_{QQ}(x_i,x_j;t) &  K_{Qq}(x_i,x_j;t) \\
\end{array}
\right)\\
&=&\label{a_sig_mat}
\left(\begin{array}{cc}
q(x_i)^T & 0 \\
Q(x_i)^T & 1 \\
\end{array}\right)
\left(\begin{array}{cc}
Z & 0 \\
0 & 1 \\
\end{array}\right)
\left(\begin{array}{cc}
-Q(x_j) & q(x_j) \\
H(x_i,x_j;t) & 0 \\
\end{array}\right)\quad.
\end{eqnarray}
Here we have used a quasi block notation where $1$ and $Z$ stand for the
$N\times N$ unit matrix or the matrix with entries $Z_{nm}$ respectively.
$q(x)$ (or $Q(x)$) is a column vector with entries $q_n^{(t)}(x)$
(or $Q_n^{(t)}(x)$).
$q(x)^T$ and $Q(x)^T$ denote the corresponding row vectors. The determinant
of (\ref{a_sig_mat}) is immediately calculated
\begin{equation}
\label{deta_calc}
\det(A(\sigma(x_i,x_j;t)))=
\det\left(q_n^{(t)}(x_i)\right)^2\,\det(H(x_i,x_j;t))\sim \left(p(x,t)\right)^2
\end{equation}
where we have expressed the determinant of the skew orthogonal polynomials
as a Vandermond determinant. Eq. (\ref{deta_calc}) completes the proof of
the property (\ref{sig_prop3}).

We can now use (\ref{sig_prop1}-
\ref{sig_prop3}) and apply the theorem 4 of Ref. \cite{dyson3}.
Then, we find directly the $m$-point correlation functions as
quaternion-determinants
\begin{equation}
\label{korr_quat}
R_m(x_1,\ldots,x_m;t)=\mbox{QDet}\left(\sigma(x_i,x_j;t)_{1\le i,j\le m}
\right)\quad.
\end{equation}
In addition, the constant in (\ref{sig_prop3}) must have the value
$1/N!$.

\subsection{Conductance correlation functions and averages.}

\label{section:2c}

A more explicit evaluation of Eq. (\ref{korr_quat})
for the $m$-point correlation functions requires the
knowledge of the skew orthogonal polynomials $q_n^{(t)}(x)$ and of
their dual functions $Q_n^{(t)}(x)$ given by (\ref{dual_def}).
The calculation of Appendix \ref{append:a} shows that
the skew orthogonal polynomials are not unique and one has to
specify a certain choice. Of course, the final results of the last
subsection do not  depend on the particular choice. A possible choice
is
\begin{eqnarray}
\nonumber
q_{2n}^{(t)}(x) & = & {\textstyle \frac{1}{2}(1+4n)}\,
P_{2n}(x)\,e^{\eps_{2n}\,t}+
s_n^{(t)}(x)\quad,\\
\label{qn_result}
q_{2n+1}^{(t)}(x) & = & {\textstyle \frac{1}{2}(3+4n)}\,
P_{2n+1}(x)\,e^{\eps_{2n+1}\,t}+
s_n^{(t)}(x)
\end{eqnarray}
where $P_n(x)$ are the Legendre polynomials and
\begin{equation}
\label{sn_def}
s_n^{(t)}(x)=\sum_{m=0}^{2n-1} (-1)^m\,
{\textstyle \frac{1}{2}(1+2m)}\,P_m(x)\,e^{\eps_m\,t}\quad.
\end{equation}
The complete calculation that yields Eq. (\ref{qn_result}) is done
in Appendix \ref{append:a}. Eq. (\ref{qn_result}) can be inverted with
the result
\begin{eqnarray}
\nonumber
{\textstyle \frac{1}{2}(1+4n)}\,P_{2n}(x)\,e^{\eps_{2n}\,t} & = &
q_{2n}^{(t)}(x) - s_n^{(t)}(x)\quad,\\
\label{qn_invert_result}
{\textstyle \frac{1}{2}(3+4n)}\,P_{2n+1}(x)\,e^{\eps_{2n+1}\,t} & = &
q_{2n+1}^{(t)}(x) - s_n^{(t)}(x)
\end{eqnarray}
where now the sum $s_n^{(t)}(x)$ is expressed in terms of the $q_n^{(t)}(x)$
\begin{equation}
\label{sn_def2}
s_n^{(t)}(x)=\sum_{m=0}^{2n-1} (-1)^m\,q_n^{(t)}(x)\quad.
\end{equation}
We consider now Eq. (\ref{qn_invert_result}) for $t=0$
and replace in the definition of $g(x,y;t)$ (compare Eq. (\ref{xgreen1}))
one of the Legendre polynomials $P_n(y)$ with the expansion in terms of the
$q_n^{(0)}(y)$. Then the integral in Eq. (\ref{h_def}) can be done because
the $q_n^{(0)}(y)$ are just the skew orthogonal polynomials of the
scalar product (\ref{skal_def0}). After some algebra with the sums
(it is useful to consider these type of operations as
matrix multiplications with suitable defined matrices) we arrive at
the expansion of the kernel $H(x_1,x_2;t)$
\begin{equation}
\label{hn_exp}
H(x_1,x_2;t)=\sum_{k,n=0}^\infty
e^{-(\eps_n+\eps_k)\,t}\,\mbox{sign}(k-n)\,P_n(x_1)\,
P_k(x_2)\quad.
\end{equation}
Our first application of (\ref{hn_exp}) is the calculation of the
dual functions $Q_n^{(t)}(x)$ that appear in the Eq. (\ref{dual_def}),
\begin{eqnarray}
\nonumber
Q_{2n}^{(t)} & = & P_{2n+1}(x)\,e^{-\eps_{2n+1}\,t}+S_n^{(t)}(x)\quad,\\
\label{dual_result}
Q_{2n+1}^{(t)} & = & -P_{2n}(x)\,e^{-\eps_{2n}\,t}+S_n^{(t)}(x)
\end{eqnarray}
where we have introduced the sum
\begin{equation}
\label{ssn_def}
S_n^{(t)}(x)=\sum_{m=2n+2}^\infty e^{-\eps_m\,t}\,P_m(x)\quad.
\end{equation}
We have now to insert the expansions (\ref{qn_result}),
(\ref{hn_exp}) and
(\ref{dual_result}) in the Eqs. (\ref{sig_def}), (\ref{korr_quat}) in
order to get the $m$-point correlation functions. In the following,
we state the result for the $1$- and the $2$- point functions.
First, we emphasize that this quaternion description,
as it is shown in the last subsection, is valid for
an even value of $N$ but the formulas below are valid for the
odd $N$ case, too. In Appendix \ref{append:b}, we discuss briefly
how the complications for odd $N$ can be treated. Second, we give
also an expression for the correlation function $R_{1,1}(x,t;y,t+\tau)$
that describes the probability density of finding $x$
at time $t$ and $y$ at time $t+\tau$. Its formal definition is
given in \cite{frahm1} and we calculated this function using a
standard method of functional derivatives described in the book of
Mehta \cite{mehta}.

 For convenience, we introduce the following three
functions
\begin{eqnarray}
\nonumber
S_N^{(t_1,t_2)}(x,y) & = & \sum_{n=0}^{N-1} {\textstyle
\frac{1}{2}(1+2n)}\,P_n(x)\,P_n(y)\,e^{\eps_n(t_1-t_2)}\\
\label{sss_def}
&& + \sum_{k=0}^{N-1} (-1)^{N-1-k}
\ {\textstyle \frac{1}{2}(1+2k)}
\,P_k(x)\,e^{\eps_k\,t_1}
\ \sum_{n=N}^\infty P_n(y)\,e^{-\eps_n\,t_2}\quad,\\
\label{ddd_def}
D_N^{(t_1,t_2)}(x,y) & = & \sum_{n,k=0}^{N-1}
\mbox{sign}(n-k)\,{\textstyle \frac{1}{4}(1+2n)(1+2k)}(-1)^{n+k}
 P_n(x)\,P_k(y)\,e^{\eps_n\,t_1+\eps_k\,t_2}\quad,\\
\label{iii_def}
I_N^{(t_1,t_2)}(x,y) & = & -\sum_{n,k=N}^\infty\mbox{sign}(n-k)
\,P_n(x)\,P_k(y)\,e^{-\eps_n\,t_1-\eps_k\,t_2}\quad.
\end{eqnarray}
Then the quaternion function (\ref{sig_def}) becomes
\begin{equation}
\label{sig_result_1}
\sigma(x,y;t)=
\left(\begin{array}{cc}
S_N^{(t,t)}(x,y) & D_N^{(t,t)}(x,y) \\
I_N^{(t,t)}(x,y) & S_N^{(t,t)}(y,x) \\
\end{array}\right)
\end{equation}
and we eventually get for the correlation functions
\begin{eqnarray}
\label{rr1_result}
R_1(x;t)&=& S_N^{(t,t)}(x,x)\quad,\\
\label{rr2_result}
R_2(x,y;t)&=& R_1(x;t)\,R_1(y;t)-S_N^{(t,t)}(x,y)\,S_N^{(t,t)}(y,x)
+D_N^{(t,t)}(x,y)\,I_N^{(t,t)}(x,y)\quad,\\
\nonumber
R_{1,1}(x,t;y;t+\tau)&=& R_1(x;t)\,R_1(y;t+\tau)-S_N^{(t,t+\tau)}(x,y)
\,S_N^{(t+\tau,t)}(y,x)\\
\label{rr11_result}
&& +D_N^{(t,t+\tau)}(x,y)\,I_N^{(t,t+\tau)}(x,y)+g(x,y;\tau)\,
S_N^{(t+\tau,t)}(y,x)\quad.
\end{eqnarray}
 For $\tau=0$ the function (\ref{rr11_result})
fulfills the identity
\begin{equation}
\label{rr11_ident}
R_{1,1}(x,t;y,t)=R_2(x,y;t)+\delta(x-y)\,R_1(x;t)
\end{equation}
where the last term accounts for the selfcorrelation of the eigenvalue
$x$. In the limit $t_1=t_2\to\infty$ in (\ref{sss_def}) only the first
sum survives and the product $D_N\cdot I_N$ vanishes. Then Eqs.
(\ref{rr1_result}-\ref{rr2_result}) equal the $1$- and $2$-point
correlation functions of the CUE case that are directly obtained
with the Legendre polynomials as orthogonal polynomials
\cite{baranger1}. The first corrections to this behavior for $t\gg 1$ are
proportional
$e^{-(\eps_N-\eps_{N-1})\,t}=e^{-8Nt}=e^{-t/t_c}$ where $t_c=1/(8N)$
is the critical time for the Brownian motion, already given in
\cite{frahm1}. Of particular interest is the density for one
channel which is just for $N=1$ the distribution of the conductance
$g=T=(1-x)/2$ in units of $2 e^2/h$. The result (\ref{rr1_result}) implies
\begin{equation}
\label{one_channel}
p(T;t)=2 R_1(1-2T;t)\Big|_{N=1}=\sum_{n=0}^\infty P_n(1-2T)\,
e^{-4n(n+1)t}\quad.
\end{equation}
The limiting cases are $p(T;0)=1/(2\sqrt{T})$ and
$\lim_{t\to\infty}p(T;t)=1$ and correspond to the results of the orthogonal
and unitary cases respectively given in Refs.
\cite{jalabert1,baranger1}, which have been found to agree with
microscopic numerical results \cite{baranger1}. Fig. 1 illustrates
the crossover between these limiting cases, where $p(T;t)$ is shown
for $t/t_c=0,\ 0.5,\ 1,\ 2,\ \infty$.

In the limit $N\to\infty$, it is possible to get a more explicit
expression for the density. It is useful to express the density in terms
of the angle variable $\varphi$, i.e. to consider
$\rho_1(\varphi;t)=2\sin(2\varphi)\,R_1(\cos(2\varphi),t)$.
The most important contributions
in the second sum of Eq. (\ref{sss_def}) come from terms with
indices near N, i.e.
$N-k\ll N$ and $n-N\ll N$. We can therefore linearize $\eps_k-\eps_n
\simeq -8N(n-k)$ and replace the Legendre Polynomials
by their asymptotic expression for large $n,k$. After some smoothing
over the strongly oscillating contributions we get the expression
\begin{equation}
\label{rho_varphi_1}
\rho_1(\varphi;t)\simeq \frac{2}{\pi}\left(
N+\sum_{l=0}^\infty e^{-(2l+1)(t/t_c)}\cos(2(2l+1)\varphi)\right)
=\frac{2}{\pi}\left(N+
\frac{\sinh(t/t_c)\,\cos(2\varphi)}{\cosh(2t/t_c)-\cos(4\varphi)}
\right)
\end{equation}
which simplifies in the limit $t\to 0$ to
\begin{equation}
\label{rho_varphi_limit}
\rho(\varphi;0)=\frac{2N}{\pi}+\frac{1}{4}\left(\delta_+(\varphi)-
\delta_+({\textstyle \frac{\pi}{2}}-\varphi)\right)
\end{equation}
where $\delta_+(\cdots)$ is the half-sided delta function that is only
defined for positive arguments. In terms of the transmission eigenvalues
$T$ the density $\tilde R_1(T;t)=2\,R_1(1-2T;t)$ reads
\begin{equation}
\label{dens_tt_1}
\tilde R_1(T;t)=\frac{1}{\pi}\frac{1}{\sqrt{T(1-T)}}
\left(N+\frac{\sinh(t/t_c)\,(1-2T)}{2(\sinh^2(t/t_2)+4T(1-T))}
\right)
\end{equation}
with the limit
\begin{equation}
\label{dens_tt_limit}
\tilde R_1(T;0)=\frac{N}{\pi}\frac{1}{\sqrt{T(1-T)}}+
\frac{1}{4}\left(\delta_+(T)-
\delta_+(1-T)\right)\quad.
\end{equation}
We see that the asymptotic density for the variable $\varphi$
is constant for the leading order in $N$. Eqs.
(\ref{rho_varphi_1}-\ref{dens_tt_limit}) contain also the next order,
responsible for the so-called weak-localization correction of $g$.
Eq. (\ref{dens_tt_limit})
confirms the result of
Ref. \cite{jalabert1}\footnote{The delta function at $T=0$
(or $\varphi=0$) was not found in \cite{jalabert1}, since the density was
expressed in terms of the variable $\lambda=1/T-1$ so that $T=0$ corresponds
to $\lambda\to\infty$, but this contribution is needed here to
conserve the normalization.} .

Some finite $N$-results for the density are given in
figures 2 and 3 (density for the transmission eigenvalue $T$ with $N=5$).
Fig. 2a and 2b compare the CUE- and COE-limits with the asymptotic density
(cp. (\ref{dens_tt_1})).
One can find strange that the COE-density at finite $N$ apparently fits
better the asymptotic density than the CUE-density. This is due
to the finite $N$-oscillations on the scale of the
``level-spacing'' which are stronger for the CUE than for the COE.
In the asymptotic results (\ref{dens_tt_1}) and (\ref{dens_tt_limit})
these oscillations are smoothed out.
The crossover density interpolates between the two limits.
Since the relative change for $N=5$ is
not so strong as for $N=1$, in Fig. 3 only the difference
$\tilde R_1(T,t)-\tilde R_1(T,\infty)$ is shown
for $t/t_c=0,\ 0.5,\ 1,\ 2,\ \infty$ on a larger scale.
In Fig. 4 a representative range for the 2-point correlation function
$\tilde R_2(T_0,T;t)$ for $N=5$, $t/t_c=10^{-6},\ 10^{-3},\ 10^{-2},
\ 1,\ \infty$ and $T_0=0.7$, $0.6\le T \le 0.8$ is illustrated. One
can very well see the effect already discussed at the end of subsection
\ref{section:2a} that the COE $\to$ CUE crossover happens at first
for very small differences $|T-T_0|$ if $t/t_c$ is small.
All the curves shown in Fig. 1-4 are obtained by a direct numerical
evaluation of the expressions (\ref{rr1_result}-\ref{one_channel}) and
(\ref{sss_def}-\ref{iii_def}).

The average of the conductance $g=\sum_j T_j=\frac{1}{2}\sum_j(1-x_j)$ and
the autocorrelation of the fluctuation $\delta g=g-\langle g\rangle$
can be expressed in terms of integrals with the densities
(\ref{rr1_result}-\ref{rr11_result}). These integrals can be
evaluated by the recursion relation of the Legendre polynomials
and their orthogonality relation with the result
\begin{eqnarray}
\label{gmean}
\langle g(t)\rangle & = & \frac{N}{2}-\frac{N}{2(2N+1)}\,e^{-8Nt}\quad,\\
\nonumber
\langle \delta g(t)\,\delta g(t+\tau) \rangle
&=& e^{-8N\tau}\,\biggl\{\frac{N^2}{4(4N^2-1)}+
\frac{N(N+1)}{4(2N+1)(2N+3)}\,e^{-8(2N+1)t}\\
\label{gauto}
&&\qquad\qquad+\frac{N(N-1)}{4(2N-1)(2N+1)}\,e^{-8(2N-1)t}\,-
\frac{N^2}{4(2N+1)^2} e^{-16Nt}\biggr\}\quad.
\end{eqnarray}
These expressions yield for $\tau=0$, $t\to\infty$ the CUE results
\begin{equation}
\label{g_cue}
\langle g\rangle_\klein{CUE}=\frac{N}{2}\quad,
\quad \langle \delta g^2\rangle_\klein{CUE}=\frac{N^2}{4(4N^2-1)}
\end{equation}
and for $\tau=t=0$ the COE expressions
\begin{equation}
\label{g_coe}
\langle g\rangle_\klein{COE}=\frac{N}{2}-\frac{N}{2(2N+1)}\quad,
\quad \langle \delta g^2\rangle_\klein{COE}=
\frac{N(N+1)^2}{(2N+1)^2(2N+3)}
\end{equation}
which are given in Ref. \cite{baranger1}.

In the limit $N\to\infty$, one has to keep constant the ratio $t/t_c=8Nt$
in order to get a non trivial crossover. Eqs. (\ref{gmean}) and
(\ref{gauto}) then become up to corrections of order $1/N$
\begin{eqnarray}
\label{gmean_inf}
\langle g(t)\rangle &=&\frac{N}{2}-\frac{1}{4}\,e^{-t/t_c}\quad,\\
\label{gauto_inf}
\langle \delta g(t)\,\delta g(t+\tau) \rangle
&=&\frac{1}{16}\,e^{-\tau/t_c}\,\left(1+e^{-2t/t_c}\right)\quad.
\end{eqnarray}
The average (\ref{gmean_inf}) can also be obtained by the
asymptotic density (\ref{rho_varphi_1}).

Fig. 5 and Fig. 6 give illustrations of the expressions (\ref{gmean}) and
(\ref{gauto}) for the average conductance and its variance
at $N=1,\ 5,\ \infty$.

\section{$S$-Brownian motion time and magnetic flux.}

\label{section:3}

 Till now, we have considered the Brownian motion ensemble
COE $\to$ CUE in terms of the parameter $t$. We suppose as in
Refs. \cite{jalabert1,baranger1} that the
scattering matrix of a chaotic quantum dot
is well described by a COE or a CUE matrix for zero or sufficiently
large magnetic flux respectively. Then the obvious question
concerning the relation between the Brownian motion time $t$
and the magnetic flux $\Phi$ arises. This section is devoted to this
question. The consideration of the magnetic flux
requires some microscopic model. Usually, such models are treated by
numerical or semiclassical means \cite{jalabert3,bohigas4,bohigas5}.
Here we use another approach that connects $S$ with
a Pandey-Mehta Hamiltonian \cite{pandey1} describing the crossover
between gaussian ensembles for the energy levels (GOE $\to$ GUE). The
crossover parameter of the latter has been related to the magnetic flux
for disordered systems \cite{dupuis1} and for ballistic cavities
\cite{bohigas5,bohigas4}.

The argumentation of Ref. \cite{bohigas4} relies essentially on
numerical calculations and that of Ref. \cite{bohigas5} on a
semiclassical approximation. In principle, the relation between the
Pandey-Mehta parameter and the magnetic flux is understood
due to these references. However, we reconsider this relation
in subsection \ref{section:3a}, using arguments
that account for both the disordered and the ballistic case. The known
result for the disordered case of Ref. \cite{dupuis1} is recovered. In
the ballistic chaotic case, we find that the relation between the
Pandey-Mehta parameter and the magnetic flux is also governed by the
quotient of two characteristic energies. In subsection \ref{section:3b}
we discuss briefly the typical field scales of a ballistic billiard.
In particular, for a very strong magnetic field the system becomes
integrable due to the appearance of Landau-levels. Finally, in subsection
\ref{section:3c}, we relate the Pandey-Mehta model for the Hamiltonian
with the $S$-Brownian motion ensemble for the scattering matrix.
 We find that, in order to
map the $S$-Brownian motion model onto this more microscopic model,
we need to assume a certain relation between the Brownian motion
time and the Pandey-Mehta parameter. This identification
gives identical expression for the average quantities in the limit
of small fluxes and large number of channels, but does not solve
remaining discrepancies for the correlation functions.

\subsection{Magnetic flux and Pandey-Mehta Hamiltonian}

\label{section:3a}

The generic random matrix ensemble that describes the
crossover from GOE $\to$ GUE is the Pandey-Mehta Hamiltonian
\cite{pandey1}
\begin{equation}
\label{pm_hamiltonian}
H_\alpha=H_\klein{GOE}(v^2)+i\alpha A(v^2)
\end{equation}
where $H_\klein{GOE}$ is a GOE matrix \cite{mehta} ($v^2$ being the
variance of the non diagonal elements) and $A$ is an {\it antisymmetric
matrix} whose independent matrix elements $A_{ij}$ are
gaussian distributed with variance $v^2$.
$\alpha\in[0,1]$ is the crossover parameter.
$\alpha=0$ corresponds to the GOE-case whereas $\alpha=1$ is the GUE-case.

 A typical one-particle Hamiltonian is of the form
\begin{equation}
\label{onepart_hamiltonian}
H=\frac{1}{2m}\left(\vec{p}-e\vec{A}\,\right)^2+V(\vec{r}\,)
\end{equation}
in $d=2$ or $d=3$ space dimensions
where $\vec{A}=\frac{1}{2}(\vec{B}\times\vec{r}\,)$ is the vector potential
in the symmetric gauge. The potential $V(\vec{r}\,)$ can model the boundary
of a chaotic ballistic cavity or the impurities of a disordered conductor
with bulk diffusion, such that the energy levels have Wigner-Dyson
GOE-statistics without magnetic field. We decompose $H$ in a real and
imaginary part, $H=H_1+i H_2$ with
$H_1=\frac{\vec{p}\,^2}{2m}+V(\vec{r}\,)+\frac{e^2}{2m}\vec{A}^2$\ ,
$H_2=-\frac{e}{m}\vec{p}\cdot\vec{A}=-\frac{e}{2m}\vec{B}\cdot\vec{L}$ and
$\vec{L}=\vec{r}\times\vec{p}$ is the angular momentum.
The contribution $\sim \vec{A}^2$ in $H_1$
modifies the potential. For the purpose of this subsection we
restrict ourselves to the case where the
magnetic field is sufficiently small and does not
change the GOE statistics for $H_1\simeq H(\vec{B}=0)$. The range of
magnetic field at which this behavior is not true corresponds to a strong
influence of the Landau Levels and the system can not be considered as
``chaotic''. A more detailed discussion of this topic and the relevant
scales of the magnetic field will be given in the next subsection.

We consider a certain energy $E_F$ and want to describe $H$ by the
Pandey-Mehta Hamiltonian (\ref{pm_hamiltonian}) in an energy range
$E\in[E_F-\Delta E,\ E_F+\Delta E]$. When $\vec{B}=0$ we assume that
$H=H_1$ with $H_\klein{GOE}(v^2)$. The variance $v^2$ is determined by the
requirement \cite{mehta} $v=\Delta_0\sqrt{\cal N}/\pi$ where
$\Delta_0\ (\ll \Delta E)$
is the level spacing of $H_1$ and ${\cal N}$ is the chosen dimension of
the GOE ensemble, assumed to be large. We do not discuss here whether the
Pandey-Mehta Hamiltonian gives an appropriate description of the
microscopic Hamiltonian $H$ at $\vec{B}\neq 0$. The numerical and
perturbative calculations of  Ref. \cite{dupuis1}  confirm that
the level correlations of a disordered ring in the diffusive region
agree with the correlation functions of $H_\alpha$ found in
\cite{pandey1}. The parameter $\alpha$ is then given by the identification
\cite{dupuis1}
\begin{equation}
\label{alpha_dupuis}
\alpha=\sqrt{\frac{\pi}{\cal N}\,\frac{E_T}{\Delta_0}}\left(2\pi
\frac{\Phi_\pzero}{\Phi_0}
\right)
\end{equation}
where $E_T=\hbar D/(2\pi R)^2$ ($D=$ diffusion constant, $R=$ ring radius)
is the Thouless energy and $\Phi_0=h/e$ is the flux quantum.
The dependence on ${\cal N}$ indicates that the GOE to GUE crossover is
properly described by the parameter $\lambda=\sqrt{\cal N}\alpha/\pi$
in the limit ${\cal N}\to\infty$ \cite{pandey1}.
The ballistic quantum billiard has
been studied in Ref. \cite{bohigas4} by numerical diagonalization.
Again the correlation functions of \cite{pandey1} are valid and the
parameter $\alpha$ is found to be proportional to a flux $\Phi$ applied
through a typical area of the system.

 We now give an alternative derivation of (\ref{alpha_dupuis})
based on a crude estimation of the variance of a
typical matrix element $(H_2)_{ij}$ in the eigenvector basis of
the real part $H_1$ of $H$.
Let $|i>$ be an eigenvector of $H_1$ with eigenvalue $\eps_i$.
For $\eps_i\simeq E_F$, we consider the expectation value
\begin{equation}
\label{expect_value}
<i|L_z(0)\,L_z(t)|i>=<i|L_z\,e^{iH_1 t/\hbar} L_z e^{-iH_1 t/\hbar} |i>=
\sum_j \left|<i|L_z|j>\right|^2\ e^{i(\eps_j-\eps_i)t/\hbar}
\end{equation}
where $L_z$ is the $z$-component of the angular momentum (assuming that
the direction of $\vec{B}$ is chosen as the $z$-axis). $L_z(t)$ is the
time dependent angular momentum with a dynamic determined by $H_1$.
We now average over the disorder or over the random shape of the
cavity and assume that the average of the matrix element
of the angular momentum is a smooth function of the energy difference
$\eps_j-\eps_i$, i.e. $\left\langle \left|<i|L_z|j>\right|^2\right
\rangle_{V}\simeq l(\eps_j-\eps_i)$. Eq. (\ref{expect_value}) then
becomes
\begin{equation}
\label{av_expect}
\left\langle L_z(0)L_z(t)\right\rangle_{Q,V}=
\frac{\hbar}{\Delta_0}\int d\omega\ l(\hbar\omega)
\ R_2\left({\textstyle \frac{\hbar\omega}{\Delta_0}}\right)
\ e^{i\omega t}\quad.
\end{equation}
    The subscript on the left side denotes a combined expectation value
from quantum mechanics and from the random parameters of $V$.
$R_2$ is just the
two point correlation function in the orthogonal case with the limit
$R_2(x)\simeq 1$ if $|x|\gg 1$. The inversion of the
Fourier transform gives
\begin{equation}
\label{fou_inv}
l(\hbar\omega)\ R_2\left({\textstyle \frac{\hbar\omega}{\Delta_0}}\right)=
\frac{\Delta_0}{2\pi\hbar}\int dt
\ \left\langle L_z(0)L_z(t)\right\rangle_{Q,V}\ e^{-i\omega t}\quad.
\end{equation}
The expectation value in the integral vanishes at a time scale
$|t|\gg \tau_L$
where $\tau_L$ is the correlation time for the angular momentum. Hence, the
integral itself vanishes as a function of $\omega$ at the scale
$|\omega|\gg 1/\tau_L$.
{}From this behavior we can already see the limitations of the GOE
description for $H_1$. The absolute squared matrix elements of an
arbitrary fixed operator with respect to the eigenvectors of a
GOE-Hamiltonian yields always a constant value independent of the particular
indices $i\neq j$ after the average, since there is no preferential
basis. Eq. (\ref{fou_inv}) implies that for $H_1$
this is only true for an energy difference $|\eps_j-\eps_i|\ll
\hbar/\tau_L=:E_\klein{corr}$. This limits the size of the energy range
that we may consider. Since we are interested by the off diagonal
matrix elements for $\Delta_0\ll |\eps_j-\eps_i|\ll E_\klein{corr}$,
we take $R_2(\frac{\hbar\omega}{\Delta_0})\simeq 1$. From
Eq. (\ref{fou_inv}) we get for the variance of the imaginary matrix elements
$(H_2)_{ij}=-\frac{eB}{2m}\,<i|L_z|j>$ of $H$
\begin{equation}
\label{variance1}
\left\langle |(H_2)_{ij}|^2\right\rangle\simeq
\frac{\Delta_0}{\pi\hbar}\left(\frac{eB}{2m}\right)^2
\tau_L\,\langle L_z^2\rangle_{Q,V}
\end{equation}
where the correlation time $\tau_L$ is defined as
$\tau_L\,\langle L_z^2\rangle_{Q,V}=\frac{1}{2}\int dt
\ \langle L_z(0)L_z(t)\rangle_{Q,V}$. Let $2R$ be the typical
diameter of the system and $v_F$ be the Fermi velocity. Furthermore,
we introduce a characteristic ballistic energy scale
$E_\klein{ball}=\hbar v_F/(2R)$ and the flux $\Phi=\pi R^2 B$ applied
through a circle of radius $R$. The expectation value
for the angular momentum is just $\langle L_z^2\rangle_{Q,V}=
\kappa\,m^2 v_F^2 R^2$ where $\kappa$ is a numerical factor
characteristic of the shape of the dot which can be evaluated in a
semiclassical way, as discussed in Appendix \ref{append:c}.

The variance (\ref{variance1}) becomes now
\begin{equation}
\label{variance2}
\left\langle |(H_2)_{ij}|^2\right\rangle\simeq
\frac{4\kappa}{\pi}\left(\frac{\Phi_\pzero}{\Phi_0}\right)^2
\frac{\Delta_0 E_\klein{ball}^2}{E_\klein{corr}}\quad.
\end{equation}
and has to be compared with the variance of the imaginary
matrix elements in (\ref{pm_hamiltonian}) that is just $\alpha^2 \Delta_0^2
{\cal N}/\pi^2$. This gives for the parameter $\alpha$
\begin{equation}
\label{alpha_result1}
\alpha=\sqrt{\frac{4\pi\kappa}{\cal N}}\frac{E_\klein{ball}}{\sqrt{\Delta_0\,
E_\klein{corr}}}\left(\frac{\Phi_\pzero}{\Phi_0}\right)\quad.
\end{equation}
This expression contains the dimensionless flux and three
typical energies $E_\klein{ball}=\hbar v_F/(2R)$,
$E_\klein{corr}=\hbar/\tau_L$ and the level spacing $\Delta_0$.

Let us first consider a disordered conductor where the electrons are
scatterered by many impurities. The angular
momentum correlation
time is then  the elastic scattering time $\tau_L=\tau_\klein{el}$
and the energy quotient becomes $E_\klein{ball}^2/(E_\klein{corr}\Delta_0)=
(2\pi)^2 E_T\,d/\Delta_0$ where $E_T$ is the Thouless energy
($E_T=\hbar D/(2\pi R)^2$ for a diffusion constant
$D=v_F^2 \tau_\klein{el}/d$). Eq. (\ref{alpha_result1}) confirms
the result (\ref{alpha_dupuis}) of Ref. \cite{dupuis1} for a ring
with $\kappa=1/d$ (Appendix \ref{append:c}).

 For a ballistic quantum billiard, $\tau_L$ equals the time of flight
$\tilde\kappa\, 2R/v_F$ ($\tilde\kappa$ is a further numerical
factor) through the cavity and $E_\klein{ball}=\tilde\kappa\, E_\klein{corr}$.
This gives
\begin{equation}
\label{alpha_result2}
\alpha=\sqrt{\frac{4\pi\kappa\,\tilde\kappa}{\cal N}\,
\frac{E_\klein{ball}}{\Delta_0}}
\left(\frac{\Phi_\pzero}{\Phi_0}\right)\quad.
\end{equation}
 Both, this result and Eq. (\ref{alpha_dupuis}) for the disordered case
depend on the energy ratio $E_C/\Delta_0$ where $\tau_C=\hbar/E_C$
is the typical time for the electron to cross the system, either by a
direct ballistic or by a diffusive motion.

 It is well known \cite{altshuler1}
that the level statistics of a disordered conductor are well described by
the GOE-statistics up to an energy scale $E_T$. Our derivation
yields a smaller energy scale $E_\klein{corr}=\hbar/\tau_\klein{el}$.
This is due to the fact that we use a criterion based on the matrix
elements of the angular momentum. If one estimates in a similar way
the matrix elements of the position operator, one recovers the
characteristic scale $E_T$, which matters if we consider
only for the level correlations.\footnote{ The considerations made here
for the disordered case are only valid
in the diffusive or metallic regime where the states are not localized.
In the localized regime the GOE statistics are no longer valid and, in
addition, the expectation value of $L_z^2$ would depend on the
localization length instead of the typical system size. }

 Such a difference does not matter for a chaotic ballistic dot
where those two energies merge to a single characteristic scale
$E_\klein{ball}=\hbar v_F/(2R)$. The energy ratio
$E_\klein{ball}/\Delta_0$ is of the order of $(k_F R)^{d-1}\gg 1$
($k_F$ is the Fermi momentum, for comparison: $E_F/\Delta_0\sim(k_F R)^d$).
The critical flux for the GOE to GUE crossover (for an energy interval of
order $\Delta_0$ ) is $\Phi_C\sim \Phi_0/(k_F R)^{(d-1)/2}\ll
\Phi_0$. In $d=2$ dimensions one gets $\sqrt{E_\klein{ball}/
\Delta_0}\sim (k_F R)^{1/2}\sim n^{1/4}$ where $n$ is the number of
energy levels below the Fermi energy in the quantum billiard.
This scaling behavior was already given in Ref. \cite{bohigas4} and
applied in Ref. \cite{pluhar}. In the
numerical simulations of Ref. \cite{bohigas3} the factor
$\sqrt{E_\klein{ball}/\Delta_0}$ is of the order of one because it
is rather difficult to increase $n$ to values where this scaling
behavior becomes important. The situation changes a bit in three
dimensions, where $\sqrt{E_\klein{ball}/\Delta_0}\sim (k_F R)\sim n^{1/3}$.
An estimation of the numerical factor that appears in
Eq. (\ref{alpha_result2}) is given in Appendix \ref{append:c}.

\subsection{Strong and weak magnetic field scales in a quantum billiard}

\label{section:3b}

We now briefly discuss the effects of stronger
magnetic fields in a quantum billiard. For simplicity we consider
the two dimensional case, where a free electron in a magnetic field
$B$ has Landau levels with energies $E_n=\hbar\omega_C(n+\frac{1}{2})$.
$n=0,1,2,\ldots$ is the number of the Landau band and
$\omega_C=(eB)/m$ is the cyclotron frequency. The eigenstates
are a combination of harmonic oscillator states (with frequency $\omega_c$)
in one direction and plane waves in the orthogonal direction.
The magnetic length $l_c=\sqrt{\hbar/(m\omega_C)}$ is the characteristic
length of the harmonic oscillator. $l_c$ is the typical width of the
states (in the oscillator direction) of the first Landau band with $n=0$.
At an arbitrary Energy $E_F$ the states have a width $\sim \sqrt{E_F/
(\hbar\omega_C)}\ l_c\sim v_F/\omega_c=R_\klein{lam}$ of the order of
the Lamor radius.

If in a quantum billiard of typical size $R$ the magnetic field is very
high so that $R_\klein{lam}\ll R$, the effect of the Landau Levels
becomes very strong.
There is a finite number of states of each Landau band (with energies
$\le E_F$) that do not feel the boundary because their typical width
$R_\klein{lam}$ is sufficiently small. The states near the boundary
(``edge states'') are raised to higher energies. The energy spectrum
near $E_F$ does not differ very much from the spectrum of the unbounded
plane and cannot be described by a GUE spectrum. The only effects are a
(very) small
breaking of the degeneracies and the appearance of the edge states.
The classical trajectories are
circles with radius $R_\klein{lam}$ that do not
see the boundary. If the magnetic field is lowered
(such that $l_c\ll R\ll R_\klein{lam}$) the fast electrons
at the Fermi energy $E_F$ feel very well the boundary and the energy
spectrum should be closer to a GUE statistics. But at low
energies $E<E_F$ the Landau levels still exist and the corresponding
edge states can have energies near $E_F$. The latter may well complicate
a simple GUE picture because the corresponding eigenfunctions
have a very small overlap with most of the states at $E_F$.
This effect disappears if the magnetic field becomes smaller such
that $R\le l_c$. Then the real GUE statistics apply.

Taking into account the effects described above and
the results of the last subsection, we distinguish four regions
for the magnetic field.

\begin{description}
\item[(i)] $\Phi/\Phi_0\leq \sqrt{\Delta_0/E_\klein{ball}}$, the
	GOE to GUE crossover regime. The level statistic of this regime
	is described by the correlation functions of Ref.
	\cite{pandey1}. If $\Phi/\Phi_0\ll \sqrt{\Delta_0/E_\klein{ball}}$,
	the behavior is near to a GOE but one must take into account that
	the level repulsion becomes immediately quadratic on
	a very short energy scale.

\item[(ii)] $\sqrt{\Delta_0/E_\klein{ball}}\ll \Phi/\Phi_0\le 1$, the
	GUE regime. The level statistics are described by a GUE.

\item[(iii)] $1\ll \Phi/\Phi_0\ll E_F/E_\klein{ball}$
	or $l_c\ll R\ll R_\klein{lam}$, the regime where first
	Landau level effects appear. The level statistic is roughly
	GUE with complications due to edge states.

\item[(iv)] $E_F/E_\klein{ball}\ll\Phi/\Phi_0$, or $R_\klein{lam}\ll R$,
	the Landau
	level regime. The behavior of the eigenvalues near $E_F$
	is very close to the spectrum of the Landau levels and thus
	regular.
\end{description}

This subdivision of the different magnetic field scales
remains essentially valid in three instead of two space dimensions. The
behavior of the regions (i) and (ii) is the same. In
the regions (iii) and (iv) the influence of the $z$-component of the
momentum becomes important. In the region (iv) every energy eigenvalue
is a sum of one Landau level of the $xy$-plane and a contribution due to
the kinetic energy in the $z$-direction. The Hamiltonian is (roughly)
a sum of
two regular independent parts. The situation in region (iii) is again
rather complicated due to a mixture of ``Poisson''- and ``GUE''-levels.

Our main concern is the crossover between the regions (i) and (ii)
where we suppose that the Hamiltonian is well described by the Pandey-Mehta
Hamiltonian (\ref{pm_hamiltonian}) with a parameter
$\alpha$ given by Eq. (\ref{alpha_result2}). In the next subsection
we discuss how $\alpha$ can be related with the Brownian motion time
$t$ of the COE $\to$ CUE crossover.

\subsection{$S$-Brownian motion time and Pandey-Mehta Hamiltonian}

\label{section:3c}

The $2N\times 2N$-scattering matrix $S$ of a physical system that is
described by an ${\cal N}\times {\cal N}$-random Hamiltonian $H$
can be expressed by
\begin{equation}
\label{sham_def}
S(H)=1-2\pi i\,W^\dagger\,G_+\,W
\end{equation}
where we have used the abbreviation
\begin{equation}
\label{gpm_def}
G_\pm=(E-H\pm i\pi\,W\,W^\dagger)^{-1}\quad.
\end{equation}
$W$ is a ${\cal N}\times 2N$ matrix that describes the coupling between
the states of the system with the scattering channels. The
expression (\ref{sham_def}) is well justified \cite{verbaarschot,iida1}
(and references therein) and is the usual starting point for
supersymmetric calculations
\cite{verbaarschot,pluhar,iida1,lewenkopf}.
It is not very difficult to see that $S(H)$ given by (\ref{sham_def})
is indeed a unitary matrix if $H$ is hermitian.
We take the limit ${\cal N}\to\infty$ but
the dimension $2N$ of the scattering matrix may remain finite (as
considered in Ref. \cite{pluhar}). When we consider the limit
$N\gg 1$, we nevertheless assume $N\ll {\cal N}$. We note $M=2N$
the dimension of $S$.

 The expression (\ref{sham_def}), relating an appropriate Hamiltonian
to $S$, is what we need for giving
a microscopic motivation of the $S$-Brownian motion model.
However, we assume for simplicity a crossover random matrix model
for the Hamiltonian, and not a real microscopic model.

Let us assume that the Hamiltonian $H$ contains a typical
parameter depending on the magnetic field. A small change of the
parameter leads to the replacement $H\to H+\Delta H$ which gives for $S$
\begin{equation}
\label{smodify}
S(H+\Delta H)=S(H)\,\exp(-i\delta X)
\end{equation}
where $\delta X$ is a hermitian matrix. Expanding Eq. (\ref{sham_def})
up to second order
we obtain after some matrix-algebra the expression
\begin{eqnarray}
\nonumber
\delta X&=& 2\pi\Big\{
W^\dagger\,G_-\,\Delta H\,G_+\,W\\
\label{deltax_def}
&&\qquad+W^\dagger\,G_-\,\Delta H\,G_+\,(E-H)\,G_-\,\Delta H\,G_+\,W
\Big\}+{\cal O}\left((\Delta H)^3\right)\quad.
\end{eqnarray}
We now assume that the Hamiltonian can be written in the form
\begin{equation}
\label{hh_form}
H(\gamma)=H_0+\gamma\,H_\klein{GUE}(v^2)
\end{equation}
where $H_0$ is some (fixed or random) part of the Hamiltonian that may
be rather general at the moment. $H_\klein{GUE}(v^2)$ is a GUE-random matrix
({\it statistically independent} of $H_0$)
whose elements are independent and normally distributed. The average and
variance of $H_\klein{GUE}(v^2)$ are characterized by
\begin{equation}
\label{hh_gue_char}
\langle H_\klein{GUE}(v^2)\rangle =0\quad,\quad
\langle H_\klein{GUE}(v^2)^\dagger\,A\,H_\klein{GUE}(v^2)\rangle =
v^2\,\mbox{tr}(A)\,{\bf 1}_{\cal N}
\end{equation}
where $A$ is an arbitrary ${\cal N}\times {\cal N}$ test-matrix. We
consider now an
infinitesimal change of the parameter $\gamma\to\gamma+\delta\gamma$.
Since the sum of two normally distributed variables is also normally
distributed, we can apply the decomposition
\begin{equation}
\label{gauss_rep}
(\gamma+\delta\gamma)\,H_\klein{GUE}(v^2)\quad\to\quad
\gamma\,H_\klein{GUE,1}(v^2)+\Delta\gamma\,H_\klein{GUE,2}(v^2)
\end{equation}
where $H_\klein{GUE,1}$ and $H_\klein{GUE,1}$ are two different
and completely independent GUE-matrices both characterized by
(\ref{hh_gue_char}). The variances are related through
\begin{equation}
\label{gamma_trans}
(\gamma+\delta\gamma)^2=\gamma^2+(\Delta\gamma)^2\quad
\Rightarrow\quad \Delta\gamma
\simeq\sqrt{2\gamma\,\delta\gamma}\quad.
\end{equation}
The replacement (\ref{gauss_rep}) is exact and does not incorporate
any approximation if the translation (\ref{gamma_trans}) is
properly used. We emphasize that this substitution of one set of
Gaussian variables with a sum of two independent sets of new
Gaussian variables works as fas as we are only interested
in simple but arbitrary averages of the type
$\langle f[S(H(\gamma))]\rangle$ (where $f$ is an
arbitrary function of the
scattering matrix) for a given value of $\gamma$. In the case of
correlations between different
$\gamma$ values, i.~e. if we want to calculate averages of the type
$\langle f_1[S(H(\gamma_1))]\,
f_2[S(H(\gamma_1+\gamma_2))]\rangle$,
the decomposition (\ref{gauss_rep}) can not be used because it is
not possible to vary $\gamma_2$ in this way without changing
$\gamma_1$. However, for the case $\gamma_1=0$
the dependence on $\gamma_2$ can be considered in the same way.
Therefore, the following argumentation concerning the relation
between the Brownian motion
time $t$ and the parameter $\gamma$ holds for averages
$\langle f[S(H(\gamma))]\rangle$ and the particular correlations
$\langle f_1[S(H(0))]\,f_2[S(H(\gamma))]\rangle$.

We identify the small matrix $\Delta H$ with
$\Delta\gamma\,H_\klein{GUE,2}$. The hermitian matrix $\delta X$
given by Eq. (\ref{deltax_def}) is now a random matrix matrix because
of $H_\klein{GUE,2}$. We average over this
GUE-matrix (denoted by $\langle \cdots\rangle_2$) and obtain with
the help of (\ref{hh_gue_char})
\begin{eqnarray}
\label{dx1_average}
\langle \delta X\rangle_2 & = & \frac{1}{2}v^2\,(\Delta\gamma)^2\,
\left(\mbox{tr}(G_+)+\mbox{tr}(G_-)\right)\,Q(H)\quad,\\
\label{dx2_average}
\langle \delta X^\dagger\,\tilde A\,\delta X\rangle_2 & = &
v^2\,(\Delta\gamma)^2\ \mbox{tr}(\tilde A\,Q(H))\,Q(H)
\end{eqnarray}
where now $\tilde A$ is an arbitrary $M\times M$-test matrix and
$Q(H)$ is the Wigner time delay matrix given by
\begin{equation}
\label{qqh_def}
Q(H)=2\pi\,W^\dagger\,G_-\,G_+\,W=-i\,S^\dagger\,
\frac{\partial S}{\partial E}\quad.
\end{equation}
We underline that the matrix elements $Q_{i,j}$ give the time delay
characterizing the scattering modes of the chaotic cavity. This physical
meaning is very useful for a more concrete and intuitive
understanding of this somewhat abstract $S$-Brownian motion ensemble.
We have to compare the averages (\ref{dx1_average}) and (\ref{dx2_average})
with the corresponding averages of the Brownian motion ``increment''
$\delta X^{(B)}$ \cite{dyson2,frahm1} given by
\begin{eqnarray}
\label{dx1_brown_average}
\langle \delta X^{(B)}\rangle_2 & = & 0 \quad,\\
\label{dx2_brown_average}
\langle (\delta X^{(B)})^\dagger\,\tilde A\,\delta X^{(B)}\rangle_2 & = &
D\,\delta t\ \mbox{tr}(\tilde A)\,{\bf 1}_M
\end{eqnarray}
where $\delta t$ is the change of the Brownian motion time $t$ and
$D$ is the diffusion constant (which is chosen as $D=4$ in the
applications of this paper and in Ref. \cite{frahm1}). From this we
can directly state:

{\em The Brownian motion model is the idealization in
which the following two conditions are fulfilled\/}
\begin{eqnarray}
\label{cond1}
\mbox{tr}(G_+)+\mbox{tr}(G_-) & = & 0\quad,\\
\label{cond2}
Q(H) & = & q_0(\gamma^2)\,{\bf 1}_M
\end{eqnarray}
{\em where $q_0(\gamma^2)$ is a number that may
depend on $\gamma^2$ but not on
the different realizations of\/} $H_\klein{GUE}$ {\em in\/} (\ref{hh_form}).
{\em If these conditions are valid the relation between the Brownian motion
time and the parameter $\gamma$ is determined by
$D\,\delta t=v^2(\Delta\gamma)^2\,q_0^2(\gamma^2)\simeq
v^2\,2\gamma\,\delta\gamma\,q_0^2(\gamma^2)$ which results in
the translation\/}
\begin{equation}
\label{time_translate}
t=\frac{v^2}{D}\int_0^{\gamma^2} dx\ q_0^2(x)\quad.
\end{equation}

 Assuming (\ref{cond2}) we can evaluate $q_0(\gamma^2)$
as the trace of the matrix $Q(H)$ which is just {\it Wigner's time delay}
\begin{eqnarray}
\nonumber
M\cdot q_0(\gamma^2)&=&\mbox{tr}(Q(H))=(-i)\mbox{tr}(G_+(2\pi i\,W
W^\dagger)G_-)\\
\label{q0_calc}
&=&(-i)(\mbox{tr}(G_-)-\mbox{tr}(G_+))
\end{eqnarray}
where the last equality holds due to (\ref{gpm_def}).
Let us assume (as in Ref. \cite{lewenkopf})
that the operator $H+i\pi WW^\dagger$ has complex eigenvalues
$E_\mu+i\Sigma_\mu$, $\Sigma_\mu > 0$ such that
\begin{equation}
\label{gpm_trace}
\mbox{tr}(G_\pm)=\sum_\mu \frac{E-E_\mu \mp i\Sigma_\mu}{
(E-E_\mu)^2+\Sigma_\mu^2}\simeq \mp i\frac{\pi}{\Delta}\quad.
\end{equation}
The second equality is an approximation valid if the typical values
of the imaginary parts satisfy $\Sigma_\mu\gg \Delta$ where $\Delta$
is the level spacing of $H$. Then, one can replace the
$\mu$-sum by an integral over $E_\mu$. This condition
is related to the limit $M\gg 1$ where
the conditions (\ref{cond1}) and (\ref{cond2}) make sense.
Actually, Eq. (\ref{gpm_trace}) yields
already the first condition (\ref{cond1}).

The second condition (\ref{cond2}) means that the time delays of
the different scattering modes are roughly identical (which looks natural for
a chaotic cavity) and do not fluctuate (in comparison with their mean
value). This second assumption is less natural, but can be justified
if the number of channels is large.\footnote{
If $H_0$ is a GOE random matrix, we can consider the deviation
$\delta Q=Q-1/M\, \mbox{tr}(Q)\,1_M$ from $Q$ with the matrix $q_0\,1_M$.
A measure for the fluctuations $\delta Q$ is just
$\mbox{tr}(\delta Q^2)=\mbox{tr}(Q^2)-1/M\,\mbox{tr}(Q)^2$.
We have calculated the average of this quantity using supersymmetric
technics in leading order of a large channel number
with the result
$\langle \mbox{tr}(\delta Q^2)\rangle = (2\pi/\Delta_0)^2/M =
{\cal O}(M^{-1})$
where $\Delta_0$ is the level spacing of $H_\klein{GUE}$.
We do not present this calculation based on a perturbative treatment
of the $\sigma$-model \cite{iida1}.
The relative variance
$\langle |\delta Q_{ij}|^2\rangle/ \langle Q_{ii} \rangle^2$ of an
arbitrary matrix element of $\delta Q_{ij}$ is
of the order ${\cal O}(M^{-1})$ thus verifying
the second condition (\ref{cond2}) in the limit $M\gg 1$. }

In the following, we only consider a
range for the parameter $\gamma$ that is small enough for having $\Delta$
determined by the part $H_0$ of the Hamiltonian (\ref{hh_form}),
i.e. $\Delta\simeq \Delta_0$ where $\Delta_0$ is the
level spacing of $H_0$. This assumption is very reasonable since
in the large ${\cal N}$ limit the range $\gamma\sim 1/\sqrt{\cal N}$
yields already the wanted crossover (if $H_0$ is a GOE matrix).
The number $q_0$ is now approximately independent of $\gamma$. Due to
(\ref{gpm_trace}) and (\ref{q0_calc}) it is given by
$q_0\simeq 2\pi/(M\,\Delta_0)$. Furthermore, since the
variance $v^2$ is related to the level spacing $\Delta_\klein{GUE}$ of
$H_\klein{GUE}$ via \cite{mehta} $v^2={\cal N}\,\Delta_\klein{GUE}^2/\pi^2$,
one gets for the Brownian motion time (\ref{time_translate})
\begin{equation}
\label{time1_result}
t=\frac{{\cal N}}{M^2}\,\frac{\Delta_\klein{GUE}^2}{\Delta_0^2}\,
\gamma^2
\end{equation}
where the diffusion constant $D=4$. Assuming the validity of the
conditions (\ref{cond1}) and (\ref{cond2}) with the approximation
(\ref{gpm_trace}),
the $S$-matrix of $H=H_0+\gamma\,H_\klein{GUE}(v^2)$ is described
by the Brownian motion ensemble
with an initial condition $S(t=0)=S(H_0)$ and a time parameter
(\ref{time1_result}).

Now, we assume that the initial Hamiltonian is a Gaussian orthogonal
random matrix $H_0=H_\klein{GOE}(v^2)$ and that $H(\gamma)=H_\klein{GOE}
(v^2)+\gamma\,H_\klein{GUE}(v^2)$. The $S$-Brownian motion ensemble
corresponds approximately to this choice with the $S$-matrix
expression (\ref{sham_def}) and the time-translation
$t=({\cal N}\gamma^2)/M^2$. The Brownian motion ensemble can describe
properly {\it averages and correlations between $\gamma=0$ and
$\gamma>0$}.\footnote{
We have still to ensure that the initial condition $\gamma=0$
corresponds to the circular orthogonal ensemble.
In general, the $S$-matrix described by Eq. (\ref{sham_def})
represents a wide class of random matrix ensembles depending
on the independent matrix elements of the symmetric $M\times M$-matrix
$W^\dagger W$ \cite{verbaarschot}.
In the case $H=H_\klein{GOE}$, one expects \cite{lewenkopf} that
there is one particular choice of $W^\dagger W$ in which $S(H)$ is exactly
distributed as a unitary COE random matrix.
This choice is characterized by the
requirement that the average $\langle S(H)\rangle=0$ vanishes
corresponding to a maximal coupling between the scattering
channels and the system \cite{verbaarschot}.
At the energy range $|E|\ll {\cal N}\Delta_\klein{GOE}$ one requires
\cite{verbaarschot,lewenkopf}
$W^\dagger W={\cal N}\Delta_\klein{GOE}/\pi^2\cdot 1_M$.
The typical value of the imaginary parts $\Sigma_\mu$ used in Eq.
(\ref{gpm_trace}) can then be estimated as
$\Sigma_\mu\simeq
1/{\cal N}\ \mbox{tr}(\pi W W^\dagger)=M\Delta_\klein{GOE}/\pi$
and the condition for the validity of (\ref{gpm_trace}) reads as
$M\gg 1$. }

 We can note that the random Hamiltonian $H(\gamma)$ is
not identical to the Pandey-Mehta Hamiltonian (\ref{pm_hamiltonian}).
However, we can again use a decomposition of Gaussian variables
which leads to the identification
$H_\alpha\to\sqrt{1-\alpha^2}\,H_\klein{GOE}(v^2) +
\sqrt{2}\,\alpha\,H_\klein{GUE}(v^2)$.
The factor $\sqrt{2}$ is due to the fact that the sum of the
real and the imaginary parts of the (off diagonal) matrix elements has a
doubled variance, i.e. $\langle ({H_\klein{GOE}}_{,ij})^2+(A_{ij})^2 \rangle
=2v^2=2\langle |{H_\klein{GUE}}_{,ij}|^2\rangle$. Since the typical
scale of the parameter
$\alpha$ is $\alpha\sim 1/\sqrt{\cal N}$, we can approximate
$H_\alpha\simeq H_\klein{GOE}(v^2) + \sqrt{2}\,\alpha\,H_\klein{GUE}(v^2)
=H(\sqrt{2}\alpha)$
and we obtain thus the relation $\gamma=\sqrt{2}\alpha$ leading to
the final relation between $t$ and $\alpha$
\begin{equation}
\label{time2_result}
t=t(\alpha)=\frac{2}{M^2}\,({\cal N}\,\alpha^2)=\frac{2}{M^2}\,
\pi^2\,\lambda^2
\end{equation}
where $\lambda$ is the crossover parameter used in Ref. \cite{pandey1}.

We emphasize that the identification of the Pandey-Mehta Hamiltonian
$H_\alpha$ with $H(\sqrt{2}\alpha)$ works only as fas as averages
are concerned. The correlations between $\alpha=0$ and $\alpha>0$ are
then {\bf not} properly described by the Brownian motion model and
indeed the critical field scale for these correlations obtained
from the Brownian motion ensemble is a factor $\sqrt{2}$ too small
if compared with semiclassical results \cite{jalabert3} and
a supersymmetric treatment of the Pandey-Mehta Hamiltonian
\cite{frahm3}.

\section{Validity and limit of the $S$-Brownian motion ensemble}

\label{section:4}

 The $S$-Brownian motion time $t$ is related to the crossover
parameter $\lambda=\sqrt{\cal N}\alpha/\pi$
of the Pandey-Mehta Hamiltonian (\ref{pm_hamiltonian}) via
\begin{equation}
\label{time_trans_end}
t(\lambda)=\frac{2\pi^2\,\lambda^2}{M^2}.
\end{equation}
This identification assumes the limit $M\gg 1$. One can express
$\lambda$ in terms of the magnetic flux: for a ballistic chaotic dot,
we found in subsection \ref{section:3a} the expression
\begin{equation}
\label{lambda_trans_end}
\lambda=\sqrt{\frac{4\kappa\,\tilde\kappa}{\pi}\,\frac{\hbar v_F}
{\Delta_0\,2R}}\,\left(\frac{\Phi_\pzero}{\Phi_0}\right)
\end{equation}
where $\kappa$ and $\tilde\kappa$ are numerical constants
characteristic for the shape of a dot of typical size $2R$
and of level spacing $\Delta_0$ without magnetic flux, $v_F$ is
the Fermi velocity.

We can now compare the result (\ref{gmean}) for the average
magneto conductance with that of Ref. \cite{pluhar}. From Eq.
(\ref{time_trans_end}) we obtain for the $S$-Brownian motion ensemble
\begin{equation}
\label{gmean_brown}
\langle g(\lambda)\rangle_\klein{BE}=
\frac{M}{4}-\frac{M}{4(M+1)}\,\exp\left(-\frac{8\,\pi^2\,\lambda^2}{M}\right)
\end{equation}
whereas the supersymmetric result of Ref. \cite{pluhar} reads
\begin{equation}
\label{gmean_supersym}
\langle g(\lambda)\rangle_\klein{SS}\simeq
\frac{M}{4}-\frac{M}{4(M+1)}\,\left(1+\frac{8(M-1)\,\pi^2\,\lambda^2}{M^2}
\right)^{-1}\quad.
\end{equation}
Eq. (\ref{gmean_supersym}) is not the exact analytical result, but
an accurate fit obtained after the numerical evaluation of three
integrals \cite{pluhar}. In leading order of $1/M$ and in the
region $\lambda^2\ll M/(8\pi^2)$,  both results become identical
whereas for larger values of $\lambda$
the difference between the Lorentzian and the Gaussian is more
important. This can be understood from the argumentation
developed in subsection \ref{section:3c}, which shows that the $S$-Brownian
motion model cannot be exactly mapped on the $S$-matrix expression
(\ref{sham_def}) used in Ref. \cite{pluhar}. The expressions
(\ref{dx1_average}) and
(\ref{dx2_average}) for the averages of the increment $\delta X$
show that in reality we have to deal with a more complicated
stochastic process than the simplified one implied by
Eqs. (\ref{dx1_brown_average}) and (\ref{dx2_brown_average}). Both
stochastic processes are only approximately in agreement for sufficiently
small Brownian motion times or values
of the symmetry breaking parameter, i.e. for $\lambda^2\ll M/8$, where
the results (\ref{gmean_brown}) and (\ref{gmean_supersym}) are consistent.

The correlations between two different values $\lambda$ and
$\lambda+\Delta\lambda$ are only properly described if $\lambda$
vanishes and $\Delta\lambda$ is sufficiently small. In addition, the
comparison with semiclassical \cite{jalabert3} and supersymmetric
\cite{frahm1} results gives a factor $\sqrt{2}$ for the relevant
$\Delta\lambda$ scale. This factor can be understood by the
difference between a symmetry breaking due to an imaginary {\it antisymmetric}
matrix or to a {\it fully hermitian} matrix (compare
subsection \ref{section:3c} and Ref. \cite{frahm3}).

In Ref. \cite{frahm1} it was shown that the Fokker-Planck equation
for the transmission eigenvalues $T_i$ can also be obtained for
a more general matrix $\delta X$ than described
by Eqs. (\ref{dx1_brown_average}) and (\ref{dx2_brown_average}).
If we assume that this is true for the $S$-matrix (\ref{sham_def})
with the Pandey-Mehta Hamiltonian, we have only to determine
correctly the relation between the $S$-Brownian motion time and
the parameter $\lambda$. Then, the expression (\ref{time2_result})
will only be an approximation valid in the range $\lambda^2\ll M/(8\pi^2)$
and one could use the supersymmetric result
(\ref{gmean_supersym}) to obtain a more appropriate expression
\begin{equation}
\label{time3_result}
t=\frac{1}{4M}\,\ln\left(1+\frac{8\,(M-1)\pi^2\,\lambda^2}{M^2}\right)\quad.
\end{equation}
  This modified relation assumes that the problem resulting from
the difference
between the $S$-Brownian motion model and the microscopic $S$-matrix
(\ref{sham_def}) can be solved by an appropriate rescaling of the Brownian
motion time.\footnote{
In fact, the relation (\ref{time3_result}) gives the correct
$\lambda$-dependence of the conductance fluctuations (in the limit $M\gg 1$)
if we apply this relation to (\ref{gauto_inf}) (for $\tau=0$)
and compare it with a perturbative treatment of
the $\sigma$-model \cite{frahm3}.}
This will work for the average conductance (suppression of the
weak localization correction), but {\it not for the conductance-conductance
correlations\/}.

 It is worth to underline that the crossover from orthogonal to
unitary symmetry happens at the critical $M$-{\it dependent} value
$\lambda_C=\sqrt{M/8}\,/\pi$ for the parameter $\lambda$. The corresponding
critical flux is given by
\begin{equation}
\label{phi_crit}
\Phi_C=\Phi_0\,\sqrt{\frac{M}{32\pi\kappa\,\tilde\kappa}\,
\frac{\Delta_0\,2R}{\hbar v_F}}\quad.
\end{equation}
 This is due to the $M$-dependence of the typical time for an
electron to stay in the dot. With increasing channel number $N=M/2$,
this typical time is shorter. Therefore the critical flux needed
to break the phase coherence between time reversed paths is
larger. The semiclassical analysis of the Ref. \cite{jalabert3} shows
that the distribution of the effective area $\Theta$ surrounded
by the electronic trajectories plays an important role. The comparison
(of the quadratic term) of the Lorentzian of \cite{jalabert3} with the
above expressions (\ref{gmean_brown}-\ref{gmean_supersym}) yields
just
\begin{equation}
\label{phi_crit_comp}
\Phi_C=\frac{1}{2}\alpha_\klein{cl}\,A\,\Phi_0
\end{equation}
where $A$ is the area of the two-dimensional cavity and $\alpha_\klein{cl}$
is the characteristic inverse area appearing in the exponential behavior
of the probability density
$p(\Theta)\sim\exp(-\alpha_\klein{cl}\,|\Theta|)$ for $\Theta$. Eqs.
(\ref{phi_crit}) and (\ref{phi_crit_comp}) yield directly
$\alpha_\klein{cl}\sim \sqrt{M}$ or $\langle |\Theta|\rangle
\sim 1/\sqrt{M}$.

 The studied $S$-Brownian motion ensemble has the decisive advantage that
many important properties
(Section \ref{section:2}) can be calculated exactly, even for
a finite dimension $M$ of the $S$-matrix. It is possible to relate
this model to the Pandey-Mehta Hamiltonian when
there is a large number of channels up to a certain approximation
and to express the Brownian motion time in terms of the parameter
$\lambda$ ( directly related to the magnetic flux). This yields
consistent results for the magneto conductance for
sufficiently small values of $\lambda\ll\lambda_C$ whereas in the
other limit the deviation between a Lorentzian and a Gaussian behavior
becomes relevant. This indicates that the $S$-matrix (\ref{sham_def})
with the Pandey-Mehta Hamiltonian (\ref{pm_hamiltonian}) is not
equivalent to the $S$-Brownian motion model. This latter model assumes
a {\it uniform and irreversible} diffusion process in the $S$-matrix
space being valid for short time scales. For longer times, a rescaling
of the time scale, as proposed in Eq. (\ref{time3_result}), improves
the validity of the model, as far as averages are concerned e. g.
average and variance of the conductance. In Ref. \cite{frahm3}, it is
however shown that even this rescaling of the Brownian motion time
is not sufficient for having the right autocorrelation function
between different times.

 To improve this exactly solvable model, one might reconsider its two
main assumptions: first that the infinitesimal evolution $\exp(-i\delta X)$
of $S$ is statistically independent of $S$ and second that this evolution
is isotropic with a time delay matrix $Q(H)$ fulfilling the conditions
Eq. (\ref{cond2}) and (\ref{cond1}). In principle, these conditions
are only valid in average. To remove the first hypothesis would imply
a Brownian motion with memory effect or a diffusion constant becoming
a random quantity on itself: i. e. a rather involved Brownian motion
in a random medium.

 It might be interesting for future works to use the $S$-Brownian
motion ensemble for a probably simpler problem: the parametric
correlations with respect to the energy. In this case, the time delay
matrix $Q(H)$ appears in a very direct way as $\delta X/\delta E$ and
just the fluctuations around its average (which we have neglected
in the last subsection) are decisive to generate the Brownian motion
dynamics. One could numerically investigate if the statistic of $Q(H)$
in a chaotic cavity is at least approximately independent of the statistic
of $S$ and obeys the isotropy condition necessary for our model.
Furthermore, one could try to formulate a non-isotropic Brownian
motion if imposed by a more realistic time delay matrix.

\section*{Acknowledgment}

\label{Acknowledgments}

 We thank O. Bohigas, F. Leyvraz, J. Rau and H. Weidenm\"uller
for useful and stimulating discussions.
This work was supported in part by an EC contract ``Quantum
dynamics of phase coherent structures''. Klaus Frahm acknowledges
the D.F.G. for a post-doctoral fellowship.

\appendix

\section{Calculation of the skew orthogonal polynomials}

\label{append:a}

In this Appendix, we sketch the calculation of the skew orthogonal
polynomials $q_n^{(t)}(x)$ for the antisymmetric scalar product
(\ref{skal_def}). First we consider the case $t=0$.
Due to $g(x,y;0)=\delta(x-y)$ and (\ref{h_def}) the scalar product
then becomes
\begin{equation}
\label{skal_def0}
<f,g>_R\,:=<f,g>^{(0)}=\frac{1}{2}\int_{-1}^1 dy_1\int_{-1}^1 dy_2
\frac{\eps(y_2-y_1)}{\sqrt{(1-y_2)(1-y_1)}}\ f(y_1)\,g(y_2)\quad.
\end{equation}
In addition, we need the conventional symmetric scalar product
given by
\begin{equation}
\label{conv_skal_def}
<f,g>_C\,=\int_{-1}^1 dy\ f(y)\,g(y)
\end{equation}
with the Legendre polynomials $P_n(x)$ as orthogonal polynomials,
i.e. $<P_n,P_m>_C=2/(1+2n)\,\delta_{nm}$.
In the following we omit the superscript $(\cdots)^{(0)}$ and denote
$q_n(x)$ the skew orthogonal polynomials (compare Eq. (\ref{skew_orth})).
In usual random matrix applications \cite{mehta} the two
type of scalar products contain the same weight function and one
can exploit a simple symmetry due to parity $x\to -x$.
The skew orthogonal polynomials are then directly related with the
conventional orthogonal polynomials and their derivatives \cite{mehta}.
The situation here is a bit more complicated and our aim
is to expand the $q_n(x)$ in terms of the Legendre polynomials.
We use therefore the ansatz
\begin{equation}
\label{q_ansatz}
q_n(x)=\sum_{m=0}^n b_{nm} {\textstyle \frac{1}{2}}(1+2m)\,P_m(x)
\end{equation}
where the $b_{nm}$ are coefficients to be determined. We note that
both scalar products are related via the following Lemma
\begin{equation}
\label{lemma}
<f,g(x)-g(-1){\textstyle \frac{1}{\sqrt{2(1-x)}}}>_C\,=
\frac{1}{2} <f,g(x)-2(1-x)\,g'(x)>_R
\end{equation}
where $x$ is the argument of the function on the right sides.
The Lemma (\ref{lemma}) is easily verified by two times partial integration of
$<f,g>_R$ (and using that $\eps(u)=\mbox{sign}(u)$). Using the expansion
\begin{equation}
\label{sqrt_expd}
\frac{1}{\sqrt{2(1-x)}}=\sum_{k=0}^\infty P_k(x)
\end{equation}
and standard identities which relate the Legendre polynomials and
their derivates we find for an arbitrary function $f(x)$ and
for $g(x)=P_m(x)$ from (\ref{lemma})
\begin{eqnarray}
\nonumber
&&<f,\,P_m-(-1)^m\sum_{k=0}^\infty P_k>_C\,=\\
\label{leg_lemma}
&&\quad=\frac{1}{2}
<f,\,(1+2m)P_m-2(2m-1)P_{m-1}+2P_{m-1}'+2P_{m-2}'>_R\quad.
\end{eqnarray}
The coefficients $b_{nm}$ can be evaluated by $b_{nm}=<q_n,P_m>_C$
which follows directly from (\ref{q_ansatz}). This identity holds
also for $n<m$ if we put in this case $b_{nm}=0$. The skew orthogonality
of the $q_n$ implies $<q_{n},r_m>_R=0$ if $r_m$ is an arbitrary polynominal
of degree $m\le n-2$ (or $m\le n$ if $n$ is an even number).
We put $f=q_n$ if $n\ge m$
and obtain from (\ref{leg_lemma}),(\ref{q_ansatz})
\begin{equation}
\label{bcoeff_eq}
b_{nm}-(-1)^m\sum_{k=0}^n b_{nk}=\frac{1}{2}
<q_n,(1+2m)P_m-2(2m-1)P_{m-1}>_R
\end{equation}
where we have used that $P_{m-1}'$ and $P_{m-2}'$ are polynomials
with a degree $\le m-2\le n-2$. In general this equation
holds for $m\le n$ but in the case of even $n$ it is valid for $m=n+1$,
too.

We consider first the case of even $n=2l$. If $m\le 2l$ the r.h.s.
of (\ref{bcoeff_eq}) vanishes and we get therefore
\begin{equation}
\label{b_eq1}
b_{2l,m}=(-1)^m\sum_{k=0}^{2l} b_{2l,k}=(-1)^m b_{2l,2l}
\end{equation}
where the second equality holds because the sum does not depend on $m$.
For $m=2l+1$ Eq. (\ref{bcoeff_eq}) gives
\begin{equation}
\label{b_eq2}
b_{2l,2l}=\sum_{k=0}^{2l} b_{2l,k}=\frac{1}{2}<q_{2l},(3+4l)P_{2l+1}>_R
=(b_{2l+1,2l+1})^{-1}\quad.
\end{equation}
The last equality is valid because the Legendre polynomials can be
expanded in terms of the $q_n$ by inversion of (\ref{q_ansatz}). The
skew orthogonality implies that for (\ref{b_eq2}) only the
first coefficient of this expansion is needed.

We exploit now (\ref{bcoeff_eq}) for the case of odd $n=2l+1$ which
is a bit more involved since three subcases $m<2l$, $m=2l$, $m=2l+1$
have to be considered yielding the following equations
\begin{eqnarray}
\label{b_eq3}
b_{2l+1,m} & = & (-1)^m s_{2l+1} \quad,\quad m<2l\quad,\\
\label{b_eq4}
b_{2l+1,2l} & = & s_{2l+1}-\frac{1}{b_{2l,2l}}\quad\\
\label{b_eq5}
b_{2l+1,2l+1} & = & -s_{2l+1}+\frac{b_{2l+1,2l}}{b_{2l+1,2l+1}\,
b_{2l,2l}}+\frac{2}{b_{2l,2l}}
\end{eqnarray}
where we have used the abbreviation
\begin{equation}
\label{sum_abbr}
s_{2l+1}=\sum_{k=0}^{2l+1} b_{2l+1,k}\quad.
\end{equation}
For (\ref{b_eq5}) the first two coefficients in the expansion of
$P_{2l+1}$ in terms of the $q_n$ were needed.

Eqs. (\ref{b_eq2}), (\ref{b_eq4}) and (\ref{b_eq5}) yield two independent
equations for the four quantities $b_{2l,2l}$, $b_{2l+1,2l}$, $b_{2l+1,2l+1}$
and $s_{2l+1}$. The other $b_{2l,m}$, $b_{2l+1,m}$ (with $m<2l$)
are determined by Eqs. (\ref{b_eq1}) and (\ref{b_eq3}). We have therefore
two free parameters that are related to the simple
transformation $q_{2l}\to A\cdot q_{2l}$ and $q_{2l+1}\to A^{-1}\cdot
q_{2l+1}+B \cdot q_{2l}$ (with arbitrary constants
$A\neq 0$, $B$) that does not change the skew orthogonality of the $q_n$.
Hence, we can choose $b_{2l,2l}=1$ and $b_{2l+1,2l}=0$ which
result in $b_{2l+1,2l+1}=1$, $s_{2l+1}=1$ and $b_{2l,m}=b_{2l+1,m}=(-1)^m$
for $m<2l$. With these values of the coefficients $b_{nm}$ the ansatz
(\ref{q_ansatz}) yields the skew orthogonal polynomials for
$<\cdots,\cdots>^{(t)}$ at the time $t=0$.

It remains the task to find the skew orthogonal polynomials for an
arbitrary time $t>0$. We define now the polynomials
\begin{equation}
\label{qn_time_def}
q_n^{(t)}(x)=\sum_{m=0}^n b_{nm}\,
{\textstyle \frac{1}{2}}(1+2m)\,P_m(x)\,e^{\eps_n\,t}
\end{equation}
where $\eps_n$ is given by $\eps_n=(1+2n)^2$ (compare Eq. (\ref{xgreen1})).
The orthogonality of the Legendre polynomials and (\ref{xgreen1})
imply directly that
\begin{equation}
\label{qn_fold}
\int_{-1}^1 dy\ g(x,y;t)\ q_n^{(t)}(y)=q_n^{(0)}(x)=q_n(x)
\end{equation}
with $q_n(x)$ given explicitly by $(\ref{q_ansatz})$ and the above stated
values of $b_{nm}$. This relation shows immediately by virtue of Eq.
(\ref{h_def}) that the $q_n^{(t)}(x)$ are indeed a set of skew orthogonal
polynomials with respect to (\ref{skal_def}). The result (\ref{qn_time_def})
is just Eq. (\ref{qn_result}) stated in Section \ref{section:2c}.

\section{Modifications in the quaternion formulation in
the case of odd dimension}

\label{append:b}
\label{oddnumber}

The quaternion formulation as described in Section \ref{section:2b}
is only valid for the case of an even Number $N$ of channels.
Here, we briefly sketch the modifications for odd $N$.
It turns out that this case can be treated in a similar way as the
even $N$ case, but nevertheless some technical complications appear.
The following treating is rather parallel to the one of Ref.
\cite{shukla1}, but adapted to the concrete situation given here.
Since for an odd dimension the determinant of an antisymmetric
matrix vanishes, Eq. (\ref{prob_pfaff}) makes of course no sense
if $N$ is odd. The method of integration over alternate variables
yields now for the integral (\ref{xsol1}) the expression
\begin{equation}
\label{prob_pfaff_odd}
p(x,t)= \propto \prod_{i>j}|x_i-x_j|\,e^{-C_N\,t}\,
\left[
\det\left(
\begin{array}{cc}
H(x_i,x_j;t) & F(x_i;t) \\
-F(x_j;t) & 0 \\
\end{array}\right)\right]^{1/2}
\end{equation}
where
\begin{equation}
\label{fff_def}
F(x;t)=\int_{-1}^1 dy\,\frac{1}{\sqrt{2(1-y)}}\,g(x,y;t)
=\sum_{n=0}^\infty P_n(x)\,e^{-\eps_n\,t}\quad.
\end{equation}
$g(x,y;t)$ is given by Eq. (\ref{xgreen1}) and the second equality
holds due to the expansion (\ref{sqrt_expd}) of the square root in
terms of Legendre polynomials. $H(x_i,x_j;t)$ stands for an $N\times N$
subblock of an $(N+1)\times(N+1)$-matrix. $F(x_i;t)$ corresponds to the
$N+1$-th column and $-F_(x_j;t)$ corresponds to the $N+1$-th row. From
Eq. (\ref{qn_result}) we get directly the identity
\begin{equation}
\label{int_id}
\int_{-1}^1 dx\ q_n^{(t)}(x)\,F(x;t)=1\quad.
\end{equation}
Therefore, it is useful to modify the skew orthogonal polynomials
by
\begin{eqnarray}
\label{qn_mod1}
\tilde q_n^{(t)}(x) & = & q_n^{(t)}(x)-q_{N-1}^{(t)}(x)\quad,
\quad n<N-1\\
\label{qn_mod2}
\tilde q_{N-1}^{(t)}(x) & = & q_{N-1}^{(t)}(x)\quad.
\end{eqnarray}
In addition, we denote with $\tilde Q_n^{(t)}(x)$ the dual functions
that correspond to the $\tilde q_n^{(t)}(x)$, i.e. they are given
by the integral (\ref{dual_def}) where the $q_n^{(t)}(x)$
are replaced by $\tilde q_n^{(t)}(x)$. Eqs. (\ref{int_id}-\ref{qn_mod2})
imply the ``orthogonality'' relation
\begin{equation}
\label{orth_mod}
\int_{-1}^1 dx\ \tilde q_n^{(t)}(x)\,F(x;t)=\delta_{n,N-1}
\quad,\quad n\le N-1\quad.
\end{equation}
Using this notations, we redefine the quaternion function $\sigma(x,y;t)$
by
\begin{equation}
\label{sig_def_odd}
\sigma(x,y;t)=\left(
\begin{array}{cc}
\sigma_{11}(x,y;t) &  \sigma_{12}(x,y;t) \\
\sigma_{21}(x,y;t) &  \sigma_{22}(x,y;t) \\
\end{array}
\right)\quad,
\end{equation}
with
\begin{eqnarray}
\nonumber
\sigma_{11}(x,y;t) & = & -K_{\tilde q\tilde Q}(x,y;t)  +\tilde q^{(t)}_{N-1}(x)
\,F(y;t)\quad,\\
\nonumber
\sigma_{12}(x,y;t) & = & K_{\tilde q\tilde q}(x,y;t) \quad,\\
\nonumber
\sigma_{21}(x,y;t) & = & H(x,y;t)-K_{\tilde Q\tilde Q}(x,y;t) + \\
\nonumber
& & \quad+ \tilde Q^{(t)}_{N-1}(x)\,F(y;t)-F(x;t)\,\tilde
Q^{(t)}_{N-1}(y) \quad,\\
\nonumber
\sigma_{22}(x,y;t) & = & K_{\tilde Q\tilde q}(x,y;t) +F(x;t)\,
\tilde q^{(t)}_{N-1}(y) \quad.
\end{eqnarray}
The function $K$ is defined as in Eq. (\ref{kk_def}) but with the
peculiarity that the sum runs over the range $n,m=0,1,\ldots,N-2$.
Eqs. (\ref{kk_prop1}-\ref{kk_prop4}) remain valid, but in (\ref{kk_prop2})
one has to replace $N$ with $N-1$. Again the three main properties
(\ref{sig_prop1}-\ref{sig_prop3}) are valid and ensure the expression
(\ref{korr_quat}) for the $m$-point correlation function. The verification
of (\ref{sig_prop1}-\ref{sig_prop2}) is a rather lengthy but
straightforward calculation using (\ref{kk_prop1}-\ref{kk_prop4}) and
(\ref{qn_mod2}). The non-trivial property is again (\ref{sig_prop3})
on which we will now concentrate. Instead of Eq. (\ref{a_sig_mat}),
we find the identity
\begin{equation}
\label{a_sig_mat_odd}
A(\sigma(x_i,x_j;t))=\lim_{\delta\to 0} M_1(\delta)\cdot M_2(\delta)
\end{equation}
where $M_{1,2}(\delta)$ stand for the two $2N\times 2N$-matrices
\begin{eqnarray}
\label{m1_def}
M_1(\delta)&=&
\left(\begin{array}{ccc}
\tilde q(x_i)^T Z  & \delta\cdot \tilde q^{(t)}_{N-1}(x_i) & 0 \\
\tilde Q(x_i)^T Z  & \quad\delta\cdot \tilde Q^{(t)}_{N-1}(x_i)
+\delta^{-1}\cdot F(x_i;t)\quad & 1 \\
\end{array}\right)\quad,\\
\nonumber
&&\phantom{a}\\
\label{m2_def}
M_2(\delta)&=&
\left(\begin{array}{cc}
-\tilde Q(x_j) & \tilde q(x_j) \\
-\delta\cdot \tilde Q^{(t)}_{N-1}(x_j)+\delta^{-1}\cdot F(x_j;t)
& \delta\cdot \tilde q^{(t)}_{N-1}(x_j) \\
H(x_i,x_j;t)-\delta^{-2}\cdot F(x_i;t)\,F(x_j;t) \quad & 0 \\
\end{array}\right)
\end{eqnarray}
with a quasi block notation similar as in Eq. (\ref{a_sig_mat}).
The left column in $M_1$ contains two $N\times (N-1)$-matrices
and the first row in $M_2$ contains two $(N-1)\times N$-matrices.
The column (row) in the middle of $M_1$ ($M_2$) is of dimension
$2N\times 1$ (or $1\times 2N$). The blocks in
the right column of $M_1$ (or in the lower row of $M_2$) are
just two square matrices of dimension $N\times N$.
The determinant of (\ref{a_sig_mat_odd}) is now evaluated as
\begin{equation}
\label{det_a_sig_odd}
\det\left(A(\sigma(x_i,x_j;t))\right)
=\lim_{\delta\to 0}
\left[\delta^2\cdot\det\left(\tilde q_n^{(t)}(x_i)\right)^2\cdot
\det\left(-H(x_i,x_j;t)+\delta^{-2}\cdot F(x_i;t)\,F(x_j;t)\right)
\right]\quad.
\end{equation}
Using the abbreviations $a_{ij}=-H(x_i,x_j;t)$, $v_i=\delta^{-1}\,F(x_i;t)$
and the identities
\begin{eqnarray}
\label{mat_prod_id1}
\left(\begin{array}{cc}
\mbox{\bf 1} & -v \\
0 & 1 \\
\end{array}\right)
\left(\begin{array}{cc}
a+vv^T & 0 \\
0 & 1 \\
\end{array}\right)
\left(\begin{array}{cc}
\mbox{\bf 1} & 0 \\
v^T & 1 \\
\end{array}\right)
&=&
\left(\begin{array}{cc}
a & -v \\
v^T & 1 \\
\end{array}\right)\quad,\\
\nonumber
&&\phantom{a}\\
\label{mat_prod_id2}
\det\left(\begin{array}{cc}
a & -v \\
v^T & 1 \\
\end{array}\right)=
\det\left(\begin{array}{cc}
a & 0 \\
v^T & 1 \\
\end{array}\right)+
\det\left(\begin{array}{cc}
a & -v \\
v^T & 0 \\
\end{array}\right)&=&
\det\left(\begin{array}{cc}
a & -v \\
v^T & 0 \\
\end{array}\right)
\end{eqnarray}
we find that the expression in Eq. (\ref{det_a_sig_odd}) become
equal (apart from the normalization constant) to the square of
the joint probability distribution (\ref{prob_pfaff_odd}).
The second equality in (\ref{mat_prod_id2}) holds because the determinant
of an antisymmetric matrix in an odd-dimensional space vanishes.
We have so far proved the last property (\ref{sig_prop3}), too.
Hence, the $m$-point functions are again given by (\ref{korr_quat}),
but now with the quaternion function (\ref{sig_def_odd}). It turns now
out that the results (\ref{sss_def}-\ref{rr11_result}) are valid for
the odd $N$-case, too.

\section{Semiclassical evaluation of the constants $\kappa$
and $\tilde\kappa$}

\label{append:c}

In subsection \ref{section:3a}, we introduced the numerical constants
$\kappa$ and $\tilde\kappa$ that depend on the dot geometry.
In this appendix, we give a semiclassical estimation of them
for some simple geometries. The constant $\kappa$ is defined
through the relation $\langle L_z^2\rangle_{Q,V} =\kappa\, m^2 v_F^2 R^2$
where the brackets denote a combined quantum mechanical and
statistical average. In the disordered case the position $\vec{r}$ and
the momentum $\vec{p}$ of the electron fluctuate on two different time
scales $\tau_T\sim R^2/D$ ($D$ is the diffusion constant) for $\vec{r}$
and $\tau_\klein{el}\ll \tau_T$ for $\vec{p}$. In the semiclassical
limit we can therefore assume that $\vec{r}$ and $\vec{p}$ are statistically
independent and the average consists of two independent integrals
of $\vec{r}$ in the particle volume and of $\vec{p}$ on the Fermi
surface $|\vec{p}\,|=m\,v_F$. For a circle of radius $R$ in $d=2$ dimensions
we obtain $\kappa=\frac{1}{4}$ and for a sphere of radius $R$ in $d=3$
dimensions the result is $\kappa=\frac{2}{15}$. One can also consider
a $d$-dimensional ring ($d=1,2,3$) on the $xy$-plane with a thickness
much smaller than $R$ but larger than the mean free path.
The integrations then yield $\kappa=\frac{1}{d}$.

In the ballistic case we need the two constants $\kappa$ and
$\tilde\kappa$ which appear only in the combination $\kappa\,\tilde\kappa$.
They are defined by
\begin{equation}
\label{kap_bal_def}
\kappa\,\tilde\kappa\,\left(\frac{2R}{v_F}\right)\,
m^2 R^2 v_F^2=\int_0^\infty dt\ \langle L_z(0)\,L_z(t)\rangle_{Q,V}
\end{equation}
where the time dependence of the angular momentum is determined by the
dynamics of the Hamiltonian without magnetic field. In Eq. (\ref{kap_bal_def})
already the time reversal invariance was exploited. Let us now consider the
classical ballistic motion of the electron in the quantum billiard. Let
$\vec{r}_i$ be the $i$-th reflection point on the surface (with
$i=0,1,2,\ldots$ and $\vec{r}_0=\vec{r}(t=0)$) and $t_i$ the corresponding
time of the reflection. The classical path is then given by
$\vec{r}(t)=\vec{r}_i+v_F\,t\,\vec{e}_i$ where $t_i\le t < t_{i+1}$ and
$\vec{e}_i$ is the unit vector in the direction of $\vec{r}_{i+1}-\vec{r}_i$.
We can now apply the semiclassical approximation to the average
\begin{equation}
\label{semi1}
\int_0^\infty dt\ \left\langle \vec{L}(0)\cdot\vec{L}(t)
\right\rangle_{Q,V}
\simeq\sum_{i=0}^\infty \left\langle (t_{i+1}-t_i)\,
\vec{L}(t_0)\cdot\vec{L}(t_i)\right\rangle_\klein{average}
\end{equation}
with
\begin{displaymath}
\vec{L}(t_i)=m v_F\frac{\vec{r}_i\times \vec{r}_{i+1}}
{|\vec{r}_{i+1}-\vec{r}_i|}\quad.
\end{displaymath}
Let us now consider a surface that is very rough on
a scale much smaller than the diameter of the cavity. In addition, we
consider again a circle in two dimensions or a sphere in three dimensions.
We assume that the surface-roughness implies that all reflection
points $\vec{r}_i$ are statistically independent and that the average
with respect to $\vec{r}_i$ is just an integration over a circle (or sphere)
with radius $R$. Then in Eq. (\ref{semi1}) only the term with $i=0$
yields a non vanishing contribution. Let $\vartheta$ be the angle between
$\vec{r}_0$ and $\vec{r}_1$. In two dimensions, Eqs. (\ref{kap_bal_def})
and (\ref{semi1}) yield after some obvious vector algebra
\def\t2{{\textstyle \frac{1}{2}\vartheta}}
\begin{equation}
\label{kap2_res}
\kappa\,\tilde\kappa=\left\langle
\sin(\t2)\,\cos^2(\t2)\right\rangle_\klein{average in
$d=2$}=\frac{2}{3\pi}\quad.
\end{equation}
In three dimensions the angular momentum has also an $x$- and a
$y$-component and we obtain
\begin{equation}
\label{kap3_res}
\kappa\,\tilde\kappa=\frac{1}{3}\left\langle
\sin(\t2)\,\cos^2(\t2)\right\rangle_\klein{average in $d=3$}=\frac{4}{45}\quad.
\end{equation}
The relation (\ref{lambda_trans_end}) has then the form
\begin{equation}
\label{lambda_trans_num}
\lambda=C_d\,\sqrt{\frac{\hbar v_F}{\Delta_0\,R}}
\,\left(\frac{\Phi_\pzero}{\Phi_0}\right)
\end{equation}
with
\begin{eqnarray}
\label{c2_value}
C_2 & = & \frac{1}{\pi}\sqrt{\,\frac{4}{3}\,}\ \simeq\ 0.3676\quad,\\
\label{c3_value}
C_3 & = & \sqrt{\,\frac{8}{45\pi}\,}\ \simeq\ 0.2379\quad.
\end{eqnarray}
The square root factor in Eq. (\ref{lambda_trans_num}) can be
calculated by standard expressions that relate the Fermi energy
and the number $n$ of electrons in the cavity with the result:
\begin{eqnarray}
\label{sqr_fac2}
\sqrt{\frac{\hbar v_F}{\Delta_0\,R}} & = & n^{1/4}\quad
\mbox{in $d=2$}\quad,\\
\label{sqr_fac3}
\sqrt{\frac{\hbar v_F}{\Delta_0\,R}} & = & (6/\pi)^{1/6}\ n^{1/3}\quad
\mbox{in $d=3$}\quad.
\end{eqnarray}
The numerical results for an asymmetric stadium
on page 176 of Ref. \cite{bohigas3} yield a relation
$\lambda=1.5\cdot(\Phi/\Phi_0)$ for the first 500 energy levels. A
comparison with (\ref{lambda_trans_num}) and (\ref{sqr_fac2}) imply
that the corresponding number of electrons is just $n\simeq 280$ which
is just in the middle of the considered energy range. The agreement
is very good if we consider the simplicity of the above used
``rough-surface'' model. The numerical values
(\ref{c2_value}-\ref{c3_value}) should therefore be considered as (rather
good) estimations for a typical quantum billiard.

\vfill\eject
\centerline{\bf FIGURE CAPTIONS}

\begin{enumerate}

\item[Fig. 1\phantom{a}] Conductance distribution $p(T;t)$
	for $N=1$ (Eq. (\ref{one_channel})) and
        $t/t_c=0,\ 0.5,\ 1,\ 2,\ \infty$.

\item[Fig. 2a] Transmission eigenvalue density $\tilde R_1(T;t)$
        for $N=5$  and $t\to\infty$ (unitary limit) and the large
        $N$-limit $N/(\pi\sqrt{T(1-T)})$ found in Ref. \cite{jalabert1}.

\item[Fig. 2b] The same as in Fig. 2a, but for $t=0$ (orthogonal limit).

\item[Fig. 3\phantom{a}] Difference $\tilde R_1(T;t)-\tilde
R_1(T;\infty)$ for $N=5$ and $t/t_c=0,\ 0.5,\ 1,\ 2,\ \infty$.
The curve with largest 	deviation corresponds to $t/t_c=0$.

\item[Fig. 4\phantom{a}] 2-point correlation function
	$\tilde R_2(T_0,T;t)$ for $T_0=0.7$ and $0.6\le T \le 0.8$
	at $N=5$ and $t/t_c=10^{-6},\ 10^{-3},\ 10^{-2},\ 1,\ \infty$.
	One can see the beginning of the COE $\to$ CUE crossover on
        small scales $|T-T_0|$ when $t/t_c \ll 1$.

\item[Fig. 5\phantom{a}]
	The weal-localization suppression
	$\langle g\rangle_\klein{CUE}-\langle g(t)\rangle$
	for $N=1,\ 5,\ \infty$ as a function of $t/t_c$.

\item[Fig. 6\phantom{a}] Conductance fluctuations
	$\langle \delta g(t)^2\rangle$ for $N=1,\ 5,\ \infty$
	as a function of $t/t_c$.

\end{enumerate}

\vfill\eject
\includegraphics{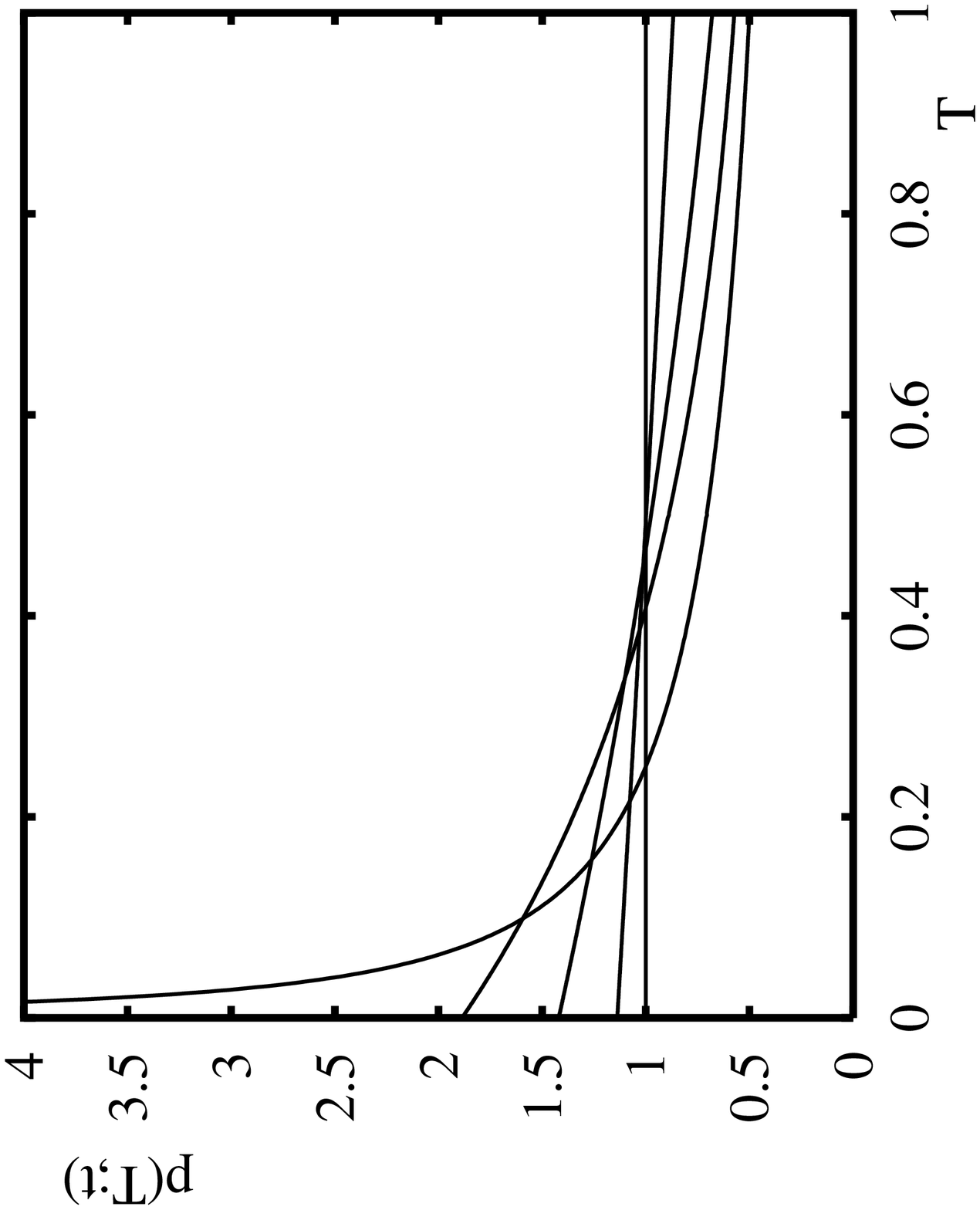}
\centerline{\LARGE Fig. 1}
\vfill\eject
\includegraphics{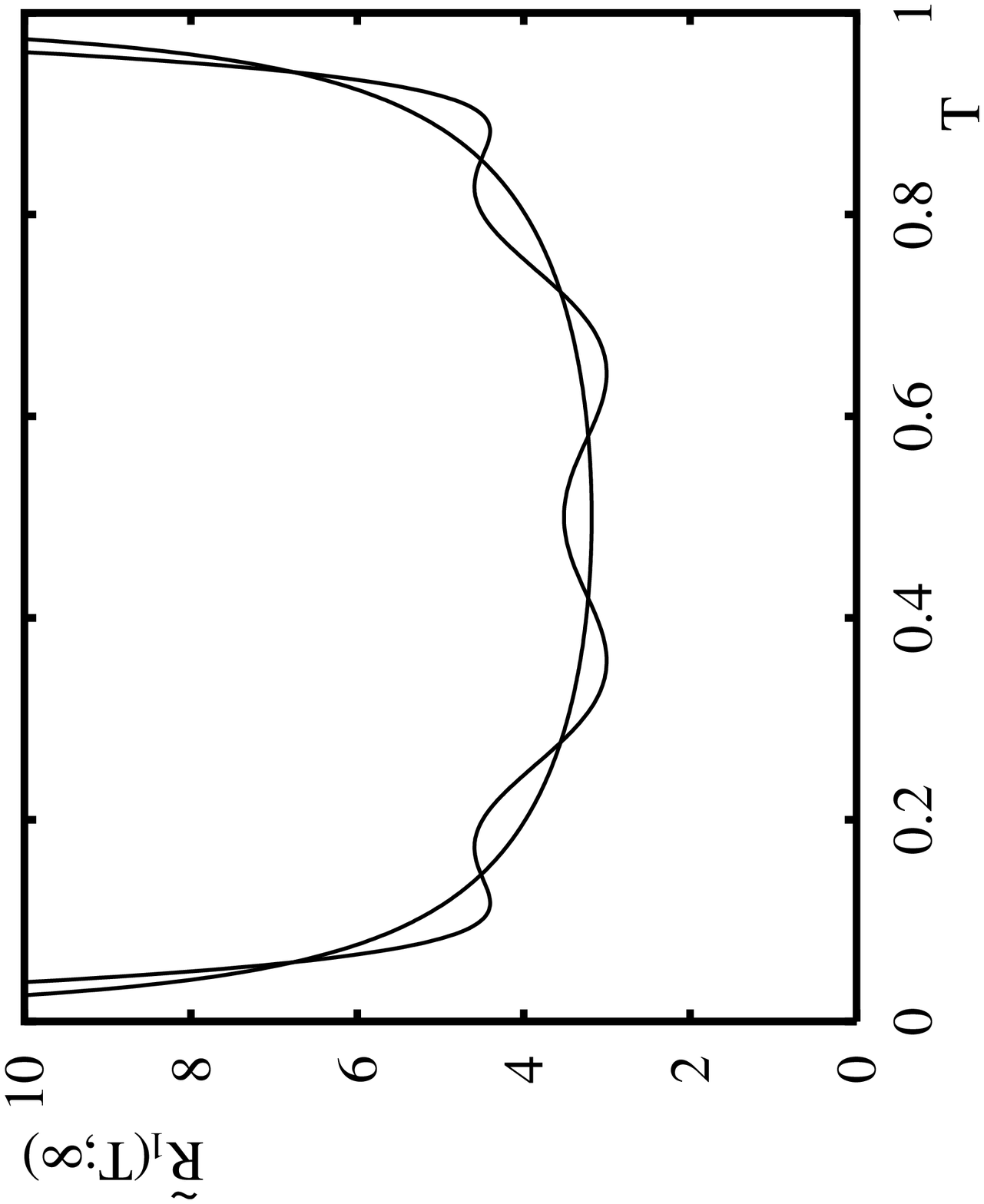}
\centerline{\LARGE Fig. 2a}
\vfill\eject
\includegraphics{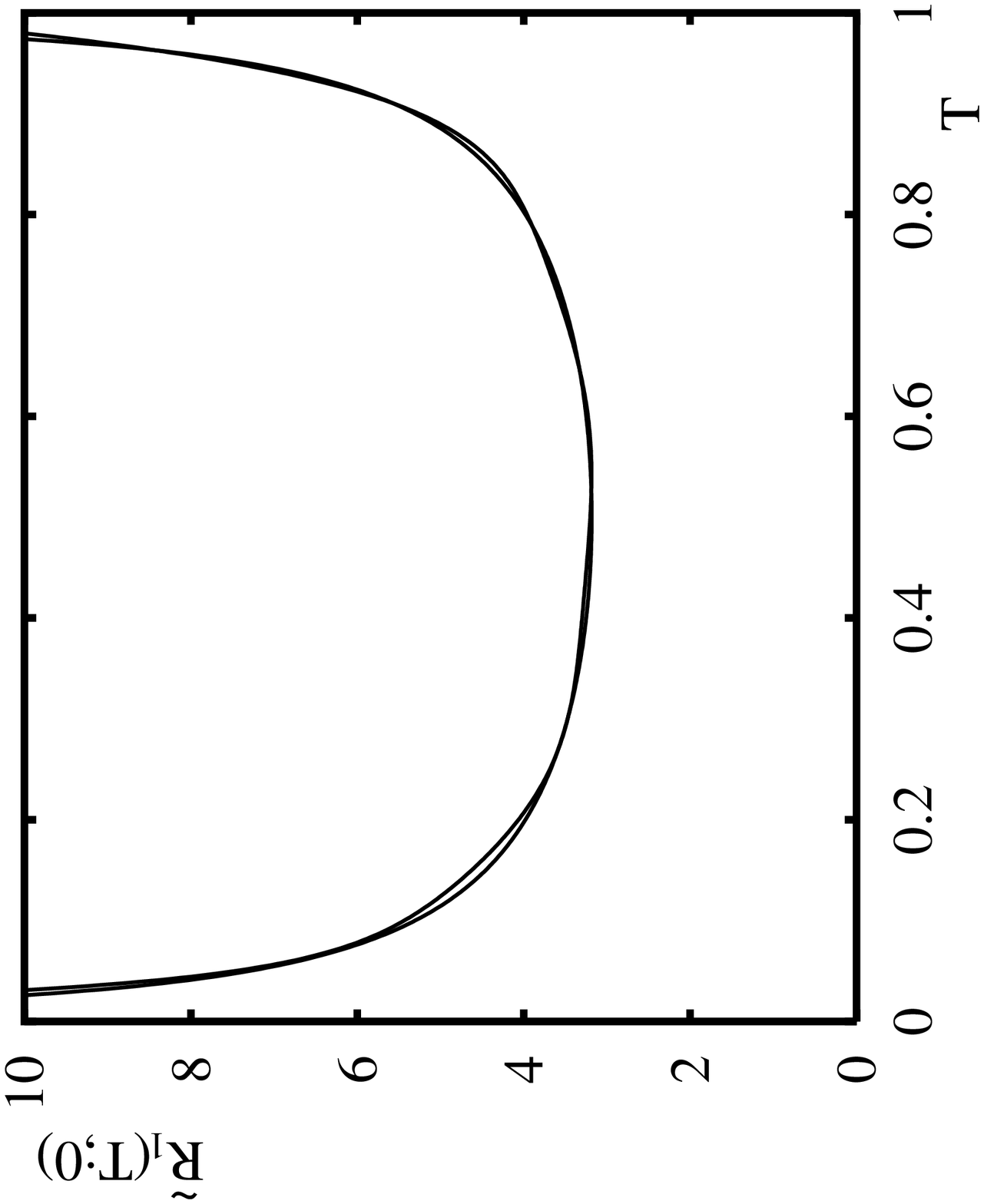}
\centerline{\LARGE Fig. 2b}
\vfill\eject
\includegraphics{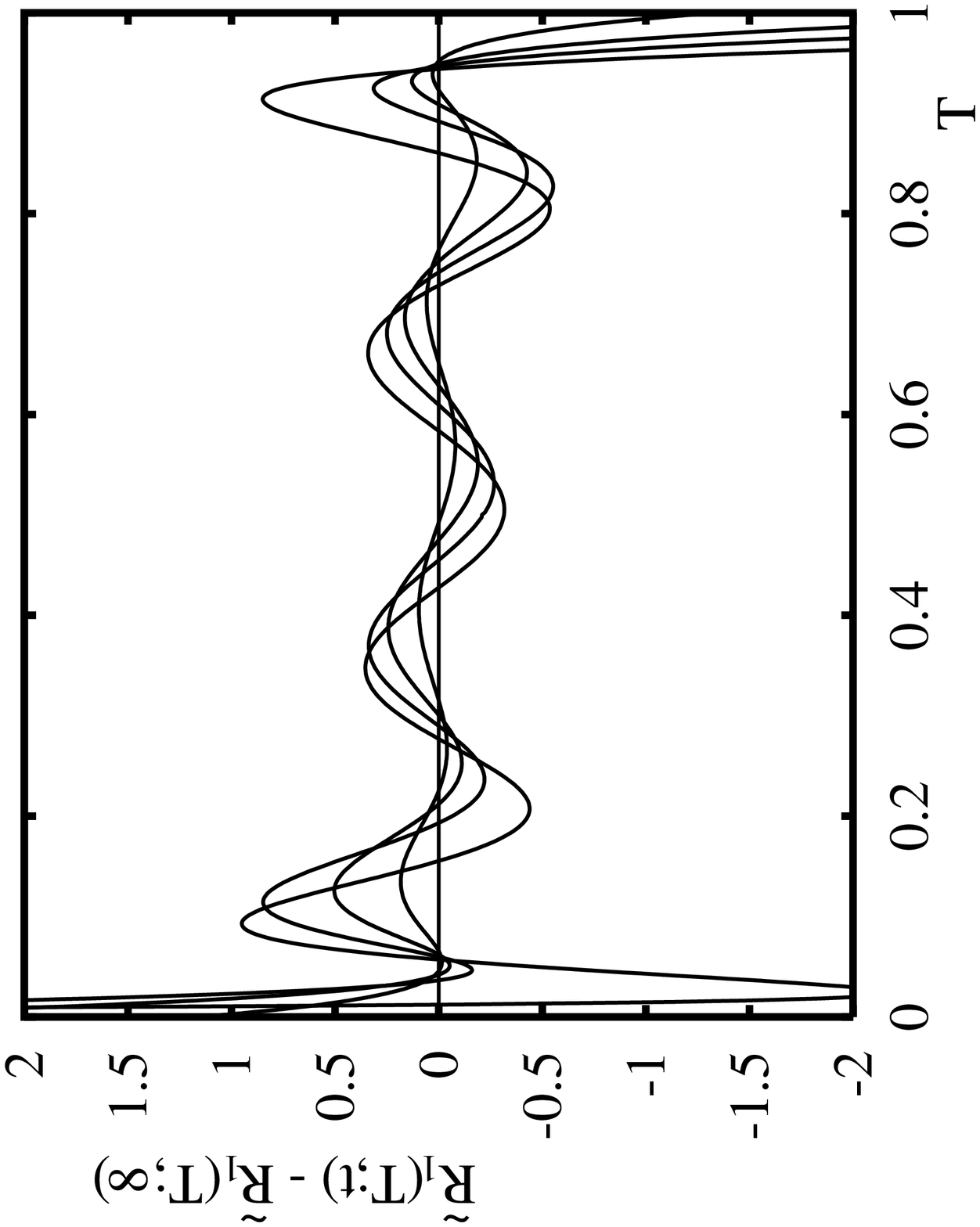}
\centerline{\LARGE Fig. 3}
\vfill\eject
\includegraphics{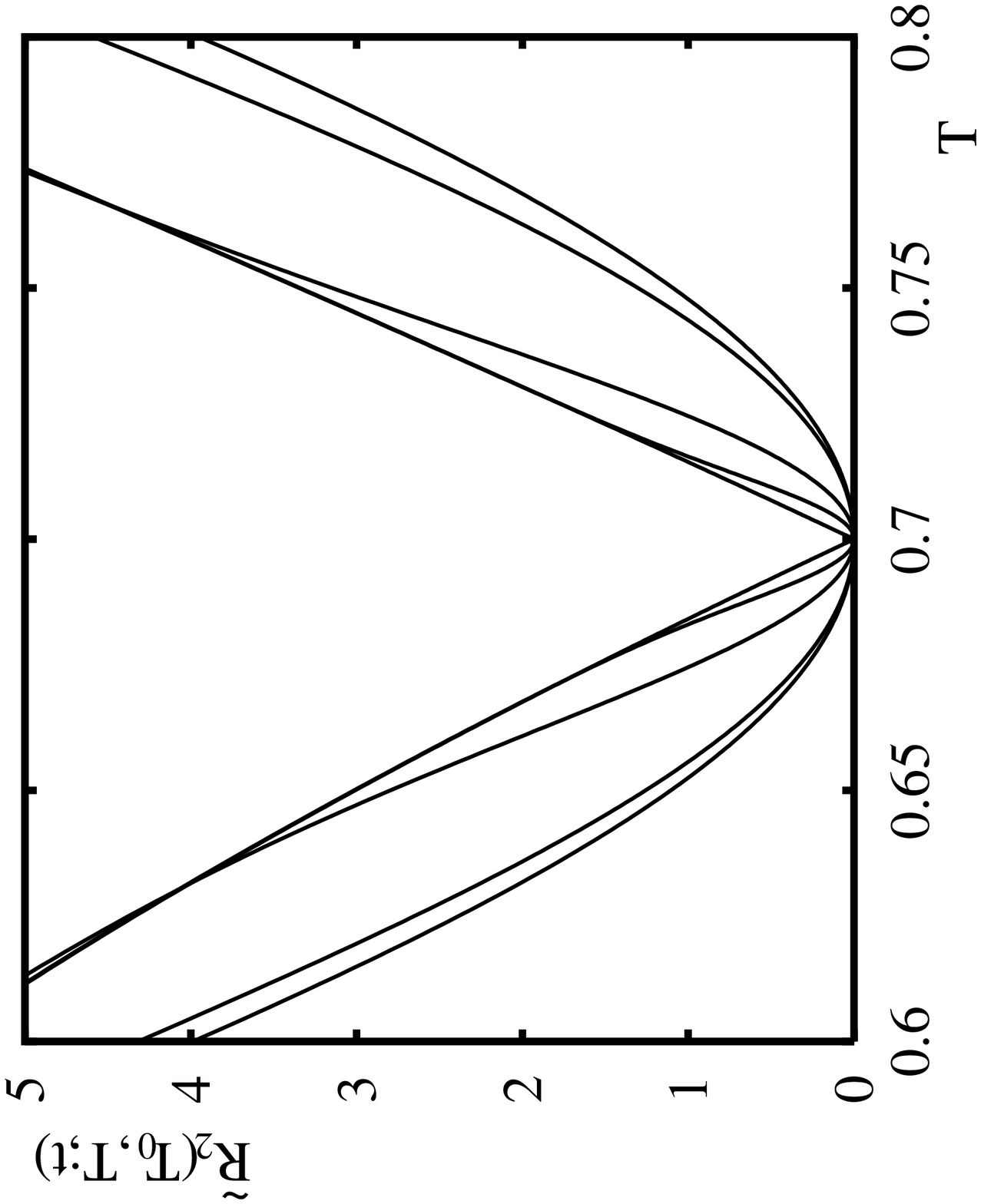}
\centerline{\LARGE Fig. 4}
\vfill\eject
\includegraphics{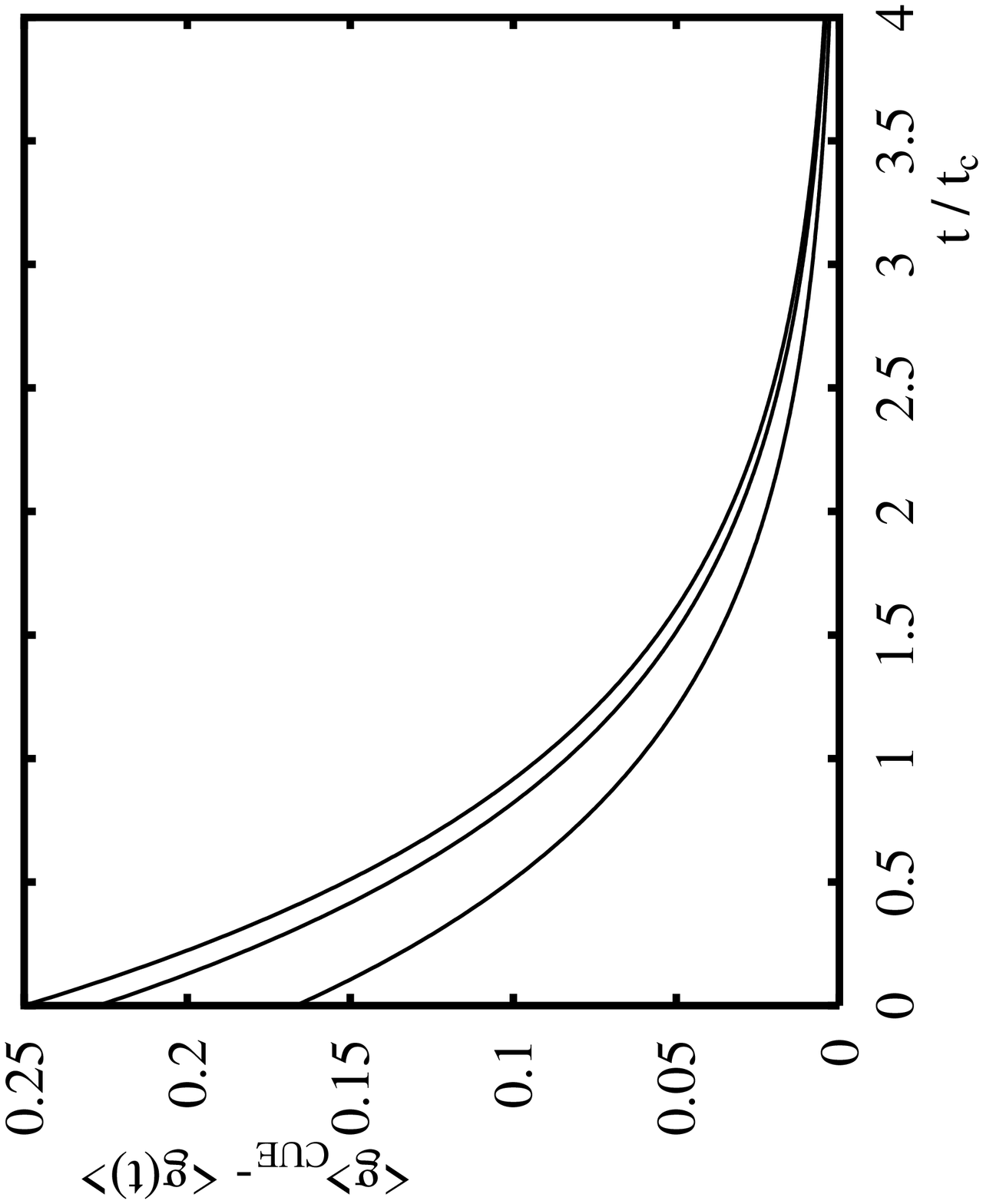}
\centerline{\LARGE Fig. 5}
\vfill\eject
\includegraphics{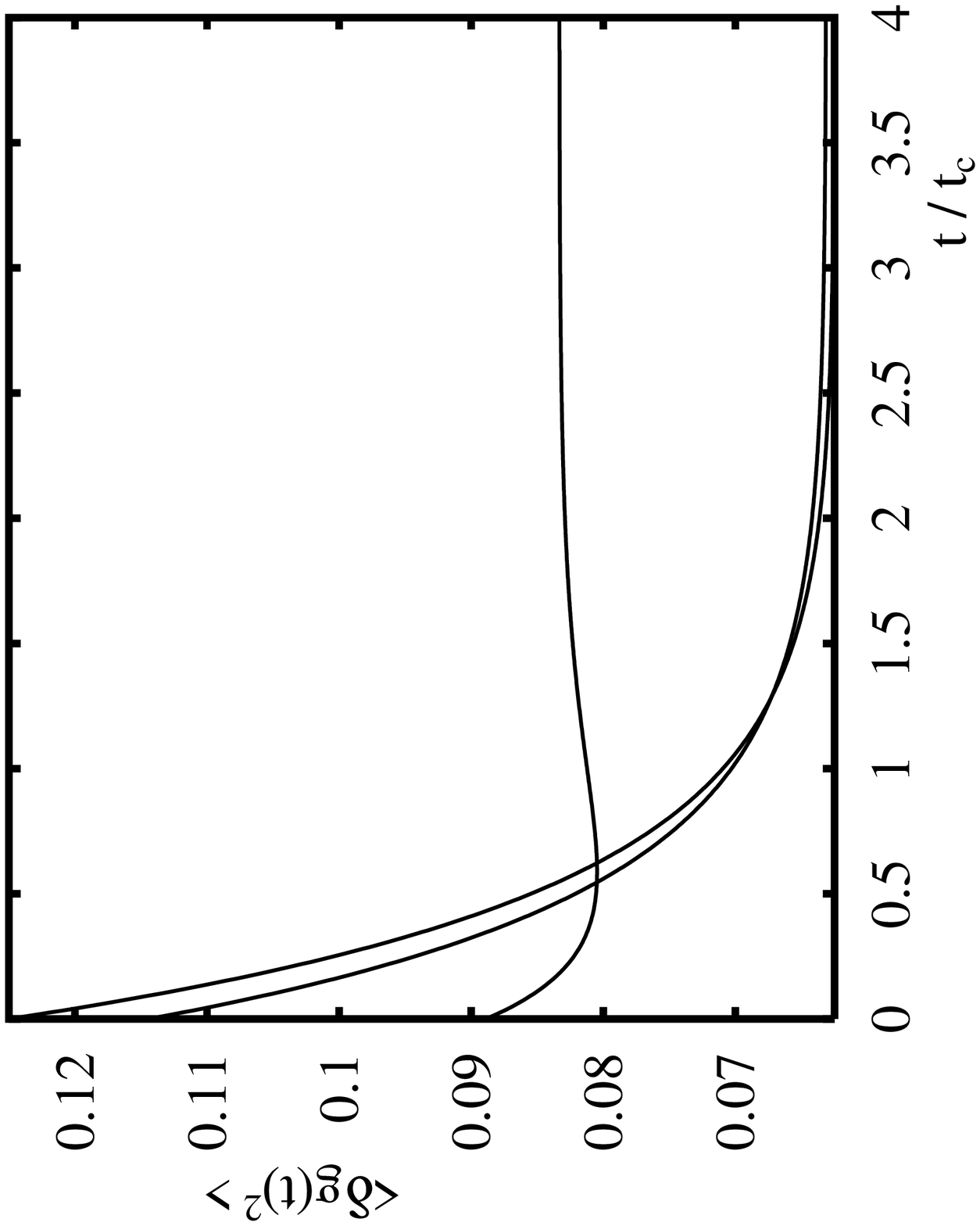}
\centerline{\LARGE Fig. 6}
\vfill\eject

\end{document}